\documentclass[11pt,preprint]{aastex}  
							
\usepackage{geometry}
\usepackage[usenames, dvipsnames]{color}
\newcommand{\Amy}{ }
\newcommand{\Revision}{ }

\usepackage{amsmath,amssymb,graphicx,fancyhdr,natbib, wasysym}

\usepackage[center,tiny]{titlesec}

\begin{document}

\section*{On the Origin of Earth's Moon}
\vspace{0.25in}
\centerline{\textbf{Amy C. Barr$^1$}}

\begin{itemize}
\item[$^1$]Planetary Science Institute, 1700 E. Ft. Lowell, Suite 106, Tucson, AZ, 85719, USA (amy@psi.edu)
\item[]Accepted for publication in Journal of Geophysical Research - Planets. Copyright 2016 American Geophysical Union. Further reproduction or electronic distribution is not permitted.
\item[]Barr, A. C. (2016), On the origin of Earth's Moon, J. Geophys. Res. Planets, 121, 1573Ð1601, doi:10.1002/2016JE005098.
\item[]*In the originally published version of this article, the target and impactor impact velocities given in Table 6 were incorrect. It listed the value for  ``Target $v_x$'' as 1.8474 km/s but has been corrected to 1.08474 km/s. The value for ``Impactor $v_x$'' was listed as -7.62 km/s but has been corrected to 7.26 km/s.  \end{itemize}

\noindent \underline{Key Points}
\begin{enumerate}
\item[1)]  The origin of the Moon is a fundamental question in planetary science.
\item[2)]  The Giant Impact is consistent with the lunar mass, angular momentum, and iron content, but not its chemical and isotopic composition.
\item[3)] Processing of lunar material through a two-phase disk yields a Moon with the proper composition, isotopic ratios, and thermal state.
\end{enumerate}

\noindent \underline{Plain Language Summary:}\\
\noindent Understanding how Earth's Moon formed is one of the most complicated and important problems in planetary science. The moon is thought to have formed when two planetary bodies collided during the late stages of planet formation. For the past few decades, scientists have been creating more sophisticated and realistic models to describe the impact, and what happens to the debris after the impact. The community has been able to piece together a self-consistent picture of the formation of the Moon that can describe the size of the Moon, its orbit, and its composition. Some details, mostly involving the chemical composition of the Moon and its similarities/differences to the Earth, need to be worked out.

\clearpage
\vspace{0.5in}
\baselineskip=15 pt
\noindent \textbf{Abstract}:  The Giant Impact is currently accepted as the leading theory for the formation of Earth's Moon.  Successful scenarios for lunar origin should be able to explain the chemical composition of the Moon (volatile content and stable isotope ratios), the Moon's initial thermal state, and the system's bulk physical and dynamical properties.  Hydrocode simulations of the formation of the Moon have long been able to match the bulk properties, but recent, more detailed work on the evolution of the protolunar disk has yielded great insight into the origin of the Moon's chemistry, and its early thermal history.  Here, I show that the community has constructed {\Amy the elements of an} end-to-end theory for lunar origin that matches the overwhelming majority of observational constraints.  In spite of the great progress made in recent years, new samples of the Moon, clarification of processes in the {\Amy impact-generated} disk, and a broader exploration of impact parameter space could yield even more insights into this fundamental and uniquely challenging geophysical problem.

\baselineskip=20pt   
\parskip=2pt	
\section{Introduction}
The origin of Earth's Moon is one of the most fundamental and vexing questions in planetary science.  {\Revision In the last few decades, the} Giant Impact theory has emerged as the leading explanation for the formation of the Moon \citep{HartmannDavis, CameronWard, Wood1986}.  The current incarnation of the theory holds that the Moon formed after a collision between two planetary embryos during the late stages of planetary accretion \citep{CanupAsphaug, Canup2004}.  Numerical hydrocode simulations have identified a specific impact scenario that seems to reproduce many features of the system, now referred to as the ``canonical impact'' \citep{Canup2014}: the young Earth experiences an oblique collision with a Mars-sized protoplanet \citep{Canup2004}, placing a little more than a lunar-mass worth of silicate-rich material into orbit.  Much of the material in the {\Revision disk} originates from the impactor \citep{Canup2004}.  In the few hundred to thousand years after the impact, the disk of material and connected silicate vapor atmosphere around the Earth cool and condense {\Revision \citep{ThompsonStevenson, Ward2011, Charnoz2015}}.  Material orbiting the Earth spreads {\Revision viscously} beyond the Roche limit, {\Revision where gravitational tidal forces are small enough to allow accretion,} permitting it to clump up and accrete into the final Moon \citep{IdaCanupStewart, Kokubo2000, SalmonCanup}.

As a geophysical problem, the origin of the Moon is uniquely challenging because there is no way to {\Revision directly} validate a numerical model of a planet-scale collision, or the evolution of a dense circumplanetary disk with laboratory data or observations.  Laboratory-scale experiments are not capable of simulating the collision of two spherical objects in the regime where their own self-gravity dominates over their strength.  {\Revision One can use laboratory experiments to construct scaling relationships between impact properties and impact outcomes, but these must be extrapolated by many orders of magnitude in scale to be applied to planetary-scale impacts.}  The planets in our Solar System are not presently arranged in orbits in which they collide; observations of collisions in other solar systems are still years away \citep{Miller-Ricci}.  Thus, it is never certain whether the simulations of the Giant Impact, or the accretion of the Moon from the impact-generated debris, are physically realistic.

In spite of this difficulty, the community has, in the last ten years, progressed from a qualitative, phenomenological understanding of processes such as the Giant Impact and the evolution of the protolunar disk, to {\Revision more quantitative, and in some cases, testable scenarios.}  Giant Impact simulations are becoming faster and easier to perform and analyze.  {\Revision Theorists are beginning to} understand how the composition of the Moon, and the mass, composition, and phase of orbiting material depends on impact conditions {\Revision \citep{Canup2001, Canup2008, NakajimaStevenson2014}.}  Only recently, {\Revision disk simulations have begun to include enough physical processes to} {\Revision provide a self-consistent picture of} how the Moon can accrete from impact-generated debris {\Revision \citep{Ward2011, SalmonCanup, Charnoz2015}}.  

The Giant Impact community is presently facing a challenge similar to the mantle convection community: how can the community reconcile the chemical/thermal history of hand samples with global-scale models?  New geochemical analyses of \emph{Apollo} samples and lunar meteorites (e.g., \citet{Touboul2007}) are providing constraints on the degree of mixing during {\Revision and after} the impact, and disk temperature/pressure histories \citep{TaylorWieczorek, Pahlevan2014} which can be used to constrain disk models.  

Here, I show that substantial progress has been made toward constructing a self-consistent, end-to-end theory for lunar origin, consistent with {\Revision many of the observed properties of the Earth and Moon.}  {\Revision These properties are discussed in Section \ref{sec:constraints}.} In Section \ref{sec:hydrocodes}, I review efforts to identify an impact scenario consistent with the mass, angular momentum, and iron fraction of the Moon.  I also discuss comparisons between numerical methods used to simulate the impact, including the hydrocodes themselves, and material models.  Material models and input parameters for a benchmarking case for Giant Impact simulations are described in the appendices of this paper.  In Section \ref{sec:chemistry}, I discuss {\Revision recent ideas about} of the structure and chemical evolution of the protolunar disk, and show that the disk could survive long enough to equilibrate isotopes of refractory species such as tungsten and titanium \citep{Zhang2012}.  In Section \ref{sec:assembly}, I show how a Moon accreted from an impact-generated disk can be consistent with the extent of melting in the young Moon.  The paper concludes with suggestions for avenues of future work, including a better understanding of disk evolution, exploration of a wider range of impact conditions, and obtaining samples of the lunar mantle.

\section{Constraints}\label{sec:constraints}
Any successful scenario for the origin of the Moon must satisfy five basic observations: 
\begin{enumerate}
\item a planet of roughly Earth's mass, $M_{E}=5.98\times10^{24}$ kg, and a Moon of mass $M_{l}=7.35\times 10^{22}$ grams,
\item a system angular momentum $L_{EM}=3.5 \times 10^{34}$ kg m$^2$ s$^{-1}$,
\item a Moon with $\sim 8\%$ iron by mass,
\item lunar stable isotope ratios similar to the terrestrial mantle, and degree of lunar volatile depletion,
\item a magma ocean initially 200-300 km thick.  
\end{enumerate}

Based solely on the partitioning of mass and angular momentum in the Earth/Moon system, it is clear that the Moon formed by a process different from the process that formed the satellites of the outer planets (Jupiter, Saturn, Uranus, and Neptune).  The Moon is anomalously large relative to the Earth, and contains much of the total angular momentum of the system \citep{MacDonald1966}.  The mass of the Moon is about 1\% the mass of the Earth, in contrast to the satellite systems of the outer planets, where the combined masses of the satellites are about $10^{-4}$ times the mass of their primaries.  Table \ref{table:ang}, reproduced from \citet{MacDonald1966} using modern values for the masses and locations of the satellites, illustrates the ratio between the angular momentum contained in the orbital motion of the satellites, $H_{sat}$, to the total system angular momentum, $H_{sat}+H_{orb}$, {\Revision where $H_{rot}$ is the spin angular momentum of the host planet}.  The Earth/Moon stands out as an anomaly: the Moon {\Revision presently contains $\sim$80\% of the total angular momentum of the Earth/Moon system.}  This value for the satellite systems of the outer planets  $\sim$1\%.  

The density of the Moon, $\bar{\rho}=3.34$ g cm$^{-3}$, is lower than the uncompressed density of the Earth, 4.4 g cm$^{-3}$, indicating a lower iron content.  The bulk lunar iron content is estimated to be 8 to 10\% \citep{Wood1986, JonesDelano, Lucey1995,  JonesPalme}, whereas the Earth is about 33\% iron by mass.  Recent re-analyses of the \emph{Apollo} seismic data confirm that the Moon has a small iron core, of radius {\Revision 325 km \citep{Williams2014}.}   

Analysis of numerous elements in lunar samples show that many stable isotopic ratios in lunar samples are nearly identical to those in the terrestrial mantle.  Oxygen isotope ratios in \emph{Apollo} {\Revision samples} lie on the same mass-dependent fractionation line as the {\Revision Earth} (e.g., \citet{Epstein1970, Wiechert2001}).   However, recent work shows that the stable isotopes of refractory species are similar as well.  Lunar anorthosites have the same $^{53}$Cr/$^{52}$Cr ratios as Earth's mantle \citep{Lugmair1998}.  Tungsten isotope ratios measured in KREEP-rich material also match those for the terrestrial mantle \citep{Touboul2007}.  Similarly, silicon isotope ratios in lunar basalts, and titanium isotopes in a wide variety of samples including whole rocks, ilmenites, and soils match terrestrial values \citep{Georg2007, Zhang2012}.  

Geochemical analyses of lunar samples also provide clues about the bulk chemical composition of the Moon, most importantly, the amount of volatile materials retained in the Moon during its accretion.  {\Revision The Moon appears to be depleted in volatile elements relative to the Earth and CI chondrites \citep{TaylorTaylorTaylor, TaylorWieczorek}.}  The pattern of element depletion provides a constraint on temperatures in the protolunar disk.  Measurements of low-Ti lunar basalts indicate depletion in elements with {\Revision solar} condensation temperatures $\lesssim 1100$ K \citep{WolfAnders1980}, including K,  Na, and Zn.  \citet{TaylorTaylorTaylor} suggest that the depletion pattern reflects a Moon accreted from high-temperature {\Revision liquid} that condensed at a temperature low enough to include K and Na, but only a very small amount of more volatile elements like Zn.  

{\Revision The bulk water content of the Moon is a topic of ongoing investigation.  Significant amounts of water, $\sim 40$ ppm, have been detected in melt inclusions of lunar volcanic glasses \citep{Saal:2008aa, Hauri2015}.  However, these materials are uncommon on the lunar surface, and could be derived from a local pocket of water-rich magmas \citep{Albarede2015}. Water contents in terrestrial and lunar apatites also appear to be similar, providing further evidence of the {\Revision flotation} of some water during lunar accretion \citep{Boyce2010, McCubbin2010, Greenwood2011,Boyce2014}.   Extrapolation of volatility curves based on measured abundances of Zn in mineral and rock fragments from \emph{Apollo} show that the bulk lunar interior may have $\leq 1$ ppm H \citep{Albarede2015}.}
 
The Moon's ancient anorthositic lunar highlands are thought to have formed from the {\Revision flotation} of buoyant plagioclase during a major melting event early in the Moon's history (see e.g., \citet{Lindy2011} for discussion).  The thickness of the crust is directly proportional to the amount of melting in the Moon \citep{Warren1985}.  Gravity data from the GRAIL spacecraft show that the crustal thickness is $\sim 34$ to 43 km {\Revision \citep{Wieczorek2013}.  GRAIL data also reveals the presence of igneous intrusions consistent with a net increase in the lunar radius of 0.6 to 4.9 km, consistent with a magma ocean 200 to 300 km thick \citep{Andrews-Hanna2013}}.  However, models of the assembly of the Moon from purely solid fragments of impact-generated debris show that the Moon accretes rapidly, on a time scale of months to years \citep{IdaCanupStewart, Kokubo2000, TakedaIda}.  A Moon formed this quickly would be unable to remove much of its accretional heat by radiation {\Revision and} conduction, and would have been completely molten \citep{PS00, CanupAnnRev2004}. 

\section{{\Revision Scenarios Constrained by Mass, Angular Momentum, and Lunar Iron Fraction} \label{sec:hydrocodes}}
Simulations of the Moon-forming impact are performed using hydrocode models, which solve conservation of mass, energy,  and momentum in a material with a given equation of state and constitutive model.  The equation of state relates pressure, density, and internal energy for each material in the domain.  A thorough description of the fundamentals of hydrocodes, the various numerical methods, and their application to various planetary problems can be found in \citet{Collins2012}.  Unlike a hydrocode used to simulate a meteorite impact, a hydrocode model of a colliding planet needs to include self-gravity, which is the dominant force on material in the domain after the shock compression has ceased.  

\subsection{Hydrocodes}
A number of modern hydrocodes have been used to simulate the Moon-forming impact.  They can be broken into two broad categories, based on the way in which the planets are represented numerically: Lagrangian approaches, and Eulerian/hybrid approaches. {\Revision In a Lagrangian hydrocode, the motions and thermodynamic evolution of individual parcels of material in the domain are tracked as a function of time.  In an Eulerian hydrocode, the domain is broken up into a mesh, and fluid motions are tracked by following flow into and out of each element in the mesh.} 

Smoothed particle hydrodynamics (SPH) is a Lagrangian method in which the impacting material is represented by parcels of equal mass, whose {\Revision volume} increases inversely proportional to their density.  A good general description of the SPH method can be found in \citet{Monaghan1992}.  Application of SPH to the Moon-forming impact in three dimensions began in the mid-1980's, with a series of five studies using increasingly realistic equations of state and higher resolution simulations to create a system with the proper mass ratio, angular momentum, and iron fraction: \citet{BenzMoonPaper1, BenzMoonPaper2, BenzMoonPaper3}, \citet{BenzMoonPaper4}, and \citet{BenzMoonPaper5}.  Searching the parameter space similar to that identified in these early papers, and using the version of SPH from \citet{BenzMoonPaper1}, \citet{CanupAsphaug}, were the first to identify the conditions of the canonical impact.

Recently, the cosmological SPH code GADGET \citep{Springel2005}, has been augmented with  equations of state relevant to planetary solids (e.g., \citet{MarcusRocky2009}).   GADGET-2 is open-source and freely available online\footnote{\url{http://wwwmpa.mpa-garching.mpg.de/gadget/}}, but one needs to implement a more sophisticated equation of state to use it for the Moon-forming impact.  This technique gives good agreement with the results of other SPH calculations, and has the benefit of being able to use large numbers of particles for more finely resolved simulations. 

Additionally, the numerical methods used in SPH continue to become more sophisticated.  For example, \citet{Hosono2016} have developed a version of SPH, ``DISPH'' in which pressure is assumed to be differentiable, rather than density.  This allows for better resolution of the core/mantle boundary and free surfaces in Giant Impact simulations.  \citet{Hosono2016} have performed initial benchmarking simulations and predict that DISPH could produce different outcomes for moon masses.  However, their simulations used a simplified equation of state, and assumed that both the protoearth and impactor were composed of pure granite, so detailed comparisons between their method and, e.g.,  the canonical impact of \citet{Canup2004}, are still pending \citep{Hosono2016}.  Two other variants of SPH involve changes in the way packets of mass are treated numerically.  One of these variants, ASPH, is presently freely distributed as Spheral++\footnote{http://sourceforge.net/projects/spheral/} \citep{OwenSpheral}.  A version of SPH has also been developed to run on GPU machines, providing {\Revision substantial} savings in computation time \citep{SchaferGPUSPH}.

Because so many of the Giant Impact simulations have been performed with SPH, it is natural to wonder whether the results can be replicated using a different numerical method.  SPH is not well-suited to computing strong shocks and shear motions that arise in an oblique impact \citep{Wada2006}.  Even though $\sim 10^{5}$ or more SPH particles may be used to represent the colliding planets, only 1\% of the particles end up in the disk, suggesting that the disk may be under-resolved \citep{Wada2006}.  To address these criticisms, \citet{Wada2006} performed the first Eulerian simulation of the Moon-forming impact with a modern hydrocode, using a modified cosmological code from \citet{WadaNorman2001}.  Unfortunately, their simulations used a simplified equations of state, which prevented a direct comparison to the modern SPH results.  

A direct comparison between SPH and Eulerian numerical methods using identical equations of state shows that the results of the canonical impact do not depend on the numerical method used.  \citet{m-moon} simulated the Moon-forming impact using CTH \citep{McGlaun}, a commercial hydrocode from Sandia National Laboratories, which solves the governing equations of shock wave propagation and deformation using an iterative approach wherein material flow is modeled first using a deformable mesh, then re-cast in a static Eulerian mesh to provide the physical and thermodynamic state for each parcel.  CTH has recently gained two crucial functionalities that allow it to model the Moon-forming impact: self-gravity and adaptive meshing (AMR) \citep{Crawford1999, Crawford2006}.  Adaptive meshing allows the element size in the numerical mesh to vary as a function of material density.  Throughout the course of the simulation, the code automatically re-meshes to place smaller elements in locations of high density (i.e., shock fronts), while placing larger elements in regions of empty space.  \citet{m-moon} find good agreement between CTH and SPH results.  Although the two simulations did not proceed identically due to differences in the extent of clumping in the disk, the final results, including the mass and composition of material orbiting the Earth, the size of the Earth, and the angular momentum partitioning between planet and disk, were {\Revision broadly similar.}    

\subsection{Equations of State}
Central to the outcome of any impact simulation is the description of the behavior of materials over the wide range of temperatures, pressures, and densities encountered during a planetary collision.  The most important of these relationships is the equation of state (EOS), which {\Revision relates pressure, density, and internal energy.}  A general discussion of equations of state for geological materials may be found in Appendix II of \citet{Melosh1989}.  

The two equations of state that have been used most widely in the Moon-forming impact are Tillotson \citep{Tillotson1962, BenzMoonPaper1, CanupAsphaug}, and ANEOS {\Revision \citep{ThompsonLauson1972, BenzMoonPaper3, Canup2004, MeloshMANEOS, m-moon}.}  The Tillotson EOS uses three different relationships between pressure and density for condensed states (where the density is greater than the uncompressed density of the material), expanded (hot) states, and intermediate densities.  One advantage of Tillotson is its relative simplicity, which makes it computationally fast.  However, Tillotson does not provide any information about the phase of the material, nor can it handle mixed phases.  This can be problematic because at a given internal energy, the pressure of the material can vary substantially depending on how much of the material is liquid versus gas \citep{BenzMoonPaper3}.  This can lead to an inaccurate calculation of pressures in mixed-phase materials, which can affect the pressure gradients that {\Revision were thought to help} inject material into orbit around the protoearth {\Revision \citep{CameronWard, StevensonSatsBook, BenzMoonPaper3}.}

A more sophisticated option is ANEOS, which describes the behavior of materials by relating the Helmholtz free energy due to atomic and electronic interactions at low temperatures, and due to thermal motion, excitation and ionization (see \citet{ThompsonLauson1972} for discussion).  ANEOS provides a thermodynamically self-consistent description of two-phase media, in which the separate phases are treated as a mixture that is in equilibrium \citep{BenzMoonPaper3}.  ANEOS coefficients for iron used by \citet{Canup2004} are summarized in Appendix \ref{sec:aneos-appendix} here.  A description of the application of ANEOS to dunite can be found in Appendix I of \citep{BenzMoonPaper3}.  

The original formulation of ANEOS did not include molecular vapor, so {\Revision an unrealistically large amount of energy was} required to vaporize solids {\Revision to atoms}, potentially leading to an underestimate of vapor production in hypervelocity impacts \citep{MeloshMANEOS}.  This prompted the development of a molecular ANEOS (M-ANEOS), which includes modifications to the inter-atomic potential at low temperatures, and changes to the Helmholtz free energy to include a partition function for molecular gasses \citep{MeloshMANEOS, m-moon}.  A detailed description of modifications to ANEOS to include molecular vapor may be found in \citet{MeloshMANEOS}.  M-ANEOS coefficients for dunite (with molecular vapor) are summarized in Appendix \ref{sec:aneos-appendix}.  

{\Revision Differences in the outcome of the Moon-forming impact depend so heavily on the equation of state (e.g., Tillotson vs. ANEOS vs. M-ANEOS) that direct comparisons between the performance of these equations of state are almost impossible.  It is entirely possible that a research group employing, e.g., Tillotson, may find that a given set of impact conditions (total mass, impact angle,  impactor-to-target mass ratio) successful, but a group employing M-ANEOS may find those conditions unsuccessful.  It is generally agreed that Tillotson, despite its convenience, has serious deficiencies and should be abandoned in favor of ANEOS or tabular equations of state.}

\subsection{Strength}
All simulations of the Moon-forming impact assume that the protoearth and impactor behave as fluids, with no strength, despite the fact that strength is shown to significantly affect the outcomes of oblique impacts in the laboratory \citep{GW1978, Schultz1996, StickleSchultz2011}.  Canonically, gravitational forces are thought to control the outcome of an impact if $\sigma_Y < \rho g R$, where $\sigma_Y$ is the yield stress of the material, $\rho$ is its density, $g$ is the local gravity, and $R$ is the length scale of the crater (in this case, one could use the impactor radius) \citep{Melosh1989}.  In the case of the Moon-forming impact, gravitational forces overwhelmingly dominate over material strength \citep{Asphaug2015}.  However, the strength of the mantles and cores of the target and impactor could affect the distribution of strain energy in each body, and consequently, the way in which the target and impactor are deformed and heated during the impact.  In an oblique impact, the shock compression is heterogeneous, leading to a non-uniform distribution of thermal energy \citep{Grady1980, KT2014}.  Localization of thermal energy along narrow zones of failure could prevent the impactor from melting completely, resulting in large-scale fragmentation and/or ricochet of the impactor \citep{SchultzGault1990, Schultz1996}, even under conditions for which analytic models may predict complete melting.  The net effect would be to decrease the effective strength of the bulk projectile, and to leave the majority of the projectile, outside of the melted regions, undisturbed \citep{PierazzoMAPS2000}.  This effect would become more pronounced as the impact becomes more grazing, leading to reduced peak shock pressures and asymmetries in shock propagation.  

In addition to projectile failure, strength could also change the way in which target material is ejected during the impact. Melting and vaporization will be localized along planes of shear failure; heterogeneities in the post-impact distribution of thermal energy will change the mix of protoearth vs. impactor material in the impact-generated disk.

\subsection{Initial Conditions}
The initial conditions of an {\Revision impact} between planetary bodies are commonly described by a ratio between the masses of the impactor and protoearth, a measure of the impact angle, and the impact velocity.  Here, I describe the descriptive quantities used by \citet{Canup2004}, but other sets of quantities (e.g., those used by \citet{StewartCuk2012}) are equally valid.  The impactor-to-total mass ratio, $\gamma=M_i/(M_i+M_t)$, where $M_i$ is the impactor mass, $M_t$ is the mass of the ``target'' (i.e., the protoearth).  The total mass involved in the impact, $M_T=M_i+M_t$.  The impact velocity, $v_{imp}$ is scaled by the mutual escape velocity of the system, $v_{esc}=\sqrt{2 G M_T/(R_i+R_t)}$, where $R_i$ and $R_t$ are the radius of the impactor and target.  For the Earth-Moon system, $v_{esc}=9.9$ km s$^{-1}$.  The impact velocity $v_{imp}^2=v_{esc}^2 + v_{\infty}^2$, where $v_{\infty}$ {\Revision represents the relative velocity of the projectile and target at large distances, before their mutual gravity accelerates them toward each other.} 

\begin{figure}[ht!]
\centerline{\includegraphics[width=60mm]{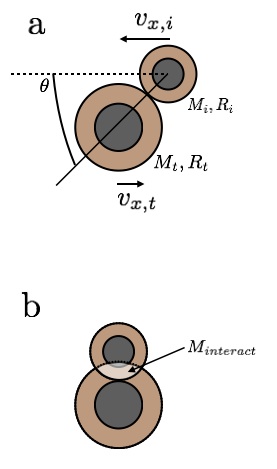}}
\caption{(a) Geometry of the impacting bodies.  Impact angle, $\theta$, is the angle between horizontal (dotted line) and the line that connects the objects' centers at the moment of impact.  A head-on impact has $\theta=0^{\circ}$, and a grazing impact has $\theta=90^{\circ}$.  Both objects have a velocity only in the $\hat{x}$ direction, calculated to keep the center of mass located close to the center of the computational domain.  The total impact velocity, $v_{imp}=|v_{x,i}|+|v_{x,t}|$ at the moment of impact.  (b) Geometrical definition of $M_{interact}$, the amount of mass contained in the lens created by the overlap of both bodies. \label{fig:geometry}}
\end{figure}

The bodies are placed in the computational domain with their centers $\sim(R_t+R_i)$ apart (see Figure \ref{fig:geometry}a), with the center of mass of the system at the center of the domain.  Each body moves toward the other in the $\hat{x}$ direction.  In an Eulerian code like CTH, it is important to adjust the velocities of each body so that the center of mass remains close to the center of the domain.  Different impact angles are obtained by changing the initial location of the impactor in the $\hat{y}$ direction.  Appendix \ref{sec:benchmarking} gives initial conditions used to simulate the canonical impact (run119 from \citet{Canup2004}, shown in Figure \ref{fig:ser119} here).  

In any hydrocode, each body must achieve hydrostatic equilibrium before the impact occurs.  In CTH, one specifies the radii of the impactor and target, the size of their iron cores, and initial guesses for the central and surface temperature of each body.  The equation of state then calculates a temperature and pressure profile consistent with hydrostatic equilibrium and the equations of state.  In SPH, one specifies the number of particles of each material, and their location, and runs SPH for a number of time steps until the impactor and target have ``settled.''  In both cases, there {\Revision are a number of possible surface and interior temperatures that yield stable initial temperature and pressure profiles.  Thus, the initial thermal state of the planets can vary.}  For ANEOS, these temperatures can be quite high, close to the {\Revision low-P melting point of ultramafic silicates} ($\sim 2000$ K).  Thus, the impactor and target may be close to the melting point before the impact even occurs.  In the absence of material strength, this may not present much of a problem, but it does raise the question of whether Moon-forming impact simulations might be overestimating melt and vapor production if both bodies are melted before the impact \citep{Canup2004}.  


\subsection{Analysis \label{sec:analysis}}
Extremely detailed descriptions of the physical behavior of the protoearth and impactor during and immediately following the Giant Impact may be found in each of the recent publications about the Moon-forming impact, including \citet{Canup2004}, \citet{StewartCuk2012}, and \citet{Canup2012}.  Here, I include a few figures showing the early evolution of the system post-impact to provide context for our discussion of disk properties and analysis of the system.  

\begin{figure}[ht!]
\centerline{\includegraphics[width=170mm]{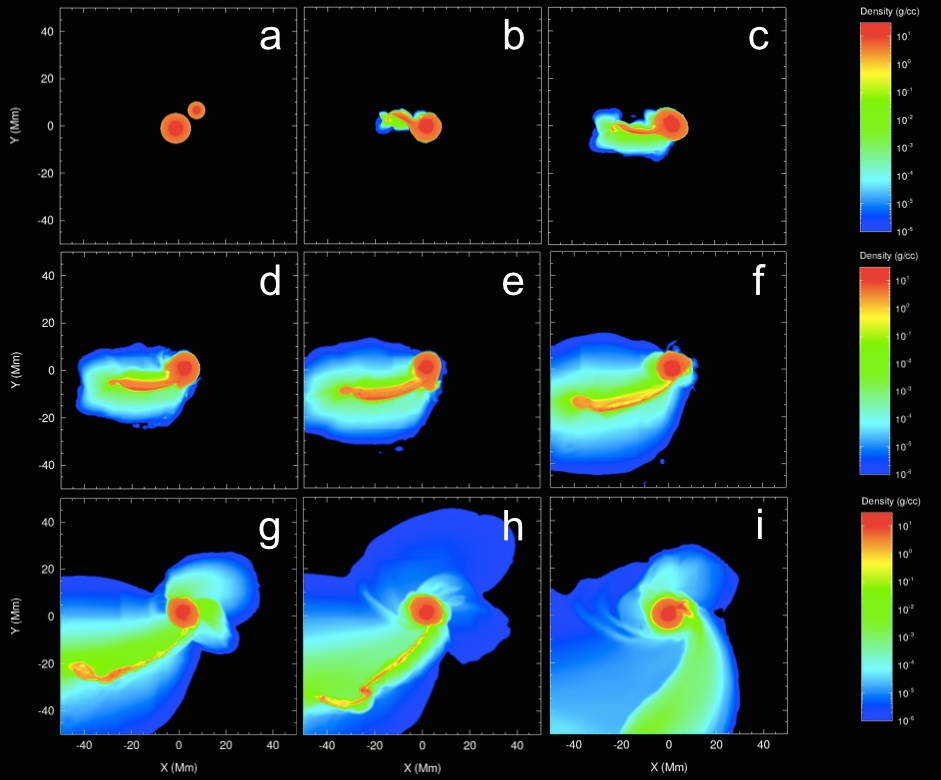}}
\caption{Snapshots of the first 10 hours of a CTH simulation \citep{m-moon} of the canonical Moon-forming impact, run119, from \citet{Canup2004}.  The impact involves a total mass $M_T=1.02M_E$, $\gamma=0.13$, $v_{imp}=v_{esc}$, an impact angle of 46$^{\circ}$, and a fully differentiated structure with 70\% dunite, 30\% iron composition by mass.  The simulation uses an ANEOS equation of state for rock and iron, with parameters listed in Appendix \ref{sec:aneos-appendix} here.  (a) $t=0$ hours, (b) $t=0.6$ hours, (c) 1 hour, (d) 1.4 hours, (e) 1.8 hours, (f) 2.4 hours, (g) 3.4 hours, (h) 4.8 hours, (i) 10 hours.  During the early stages of the impact, the impactor is sheared out into an ``arm'' of material, and the Earth is significantly deformed , including the formation of a bulge (visible at 6 o'clock in panel d) that torques the Earth, lofting material into circumplanetary orbit.  An animated version of this figure appears as a supplementary movie online.  \label{fig:ser119}}
\end{figure}

Figure \ref{fig:ser119} illustrates the first 10 hours of one of the successful canonical impacts from \citet{Canup2004}, ``run119,'' simulated with CTH \citep{m-moon}.  {\Revision \citet{Canup2004} describes the evolution of the system in the first hours after the impact: ``After about 50 minutes, the highly distorted form of the impactor extends to a distance of about 3 to 3.5 Earth radii ... and the target and inner portions of the impactor begin to rotate ahead of the distant portions of the impactor.  Both the inner orbiting impactor material and the wave/bulge on the surface of the protoearth that forms after about 80 minutes ... lead ahead of the outer portions of the impactor, providing a positive torque; after about 2 hours ... the latter bulge has propagated about two-thirds of the way around the planet from the initial impact site, while the most distant portions of the impactor, now at about 6 Earth radii from the planet's center, begin to gravitationally self-contract.''}

As demonstrated by Figure \ref{fig:ser119}, the end result of a Giant Impact simulation is a cloud of material orbiting a central concentration of mass.  It is not immediately obvious where the planet ends and the ``disk'' begins.  \citet{Canup2001} suggest a procedure to determine whether each parcel of material is part of the planet, or part of the disk, which can work for CTH and SPH data.  One picks an initial guess for the physical size of the planet, $R_{pl}$, which is likely larger than the radius of the planet pre-impact ($R_t$), because some of the impactor material has merged with the planet.  For each parcel that is  ``outside'' the planet, the parcel's location and velocity are used to calculate the radius of its equivalent circular orbit about the planet's center ($r_{circ}$) \citep{Canup2001}.  The equivalent circular radius is defined using $r_{circ}=(h_z^2/GM_{pl})$, where $M_{pl}$ is the mass of the planet, $h_z$ is the angular momentum of the parcel per unit mass, $h_z=x v_y - y v_x$, where $x$ and $y$ are the locations of the parcel relative to the center of mass of the debris cloud, and $v_x$ and $v_y$ are its velocity relative to the center of mass.  If a parcel's value of $r_{circ}$ is less than $R_{pl}$, its mass is added to the planet.  This procedure is repeated iteratively until the mass and radius of the planet converge \citep{Canup2001}.  {\Revision Since most of the bound mass is contained within the planet, once} the mass of the planet has been determined, one can determine how much material is gravitationally unbound from the system, $M_{esc}$.  

Typically, the disk mass ($M_D$), disk angular momentum ($L_D$), and iron fraction in the disk, ($m_{fe,disk}$) are tracked as a function of time.  Figure \ref{fig:ser119-timelogs} illustrates the evolution of these quantities for the simulation depicted in Figure \ref{fig:ser119}.  The simulation is considered ``complete'' when $M_D$ and $L_D$ do not change significantly between time steps.  At late simulation times, artificial viscosity dominates the momentum transfer in a shock physics code, which gives rise to a linear decrease $M_D$ and $L_D$, the slope of which depends on the viscosity parameter (see Appendix B of \citet{m-moon} for discussion).  

It is also common to report the estimated lunar mass, $M_L$.  As I will discuss in Section \ref{sec:chemistry}, determining the mass of moon (or moons) created from an impact-generated disk requires careful simulation of the evolution of the multi-phase disk inside the Roche limit, coupled with an $N$-body simulation to account for gravitational and collisional evolution of the debris.  Results of this type of modeling suggest that the predicted lunar mass depends on the ratio between the disk angular momentum per unit mass and the angular momentum per unit mass of a circular orbit with semi-major axis equal to the Roche limit around the Earth, $a_R$, and the amount of disk material that escapes, $M_{esc}$, \citep{IdaCanupStewart, SalmonCanup},
\begin{equation}
\frac{M_L}{M_D} \approx a \bigg(\frac{L_D}{M_D \sqrt{GM_{E} a_R}}\bigg) - b -c \bigg(\frac{M_{esc}}{M_D}\bigg).  \label{eq:ICS97}
\end{equation}
{\Revision where $a$, $b$, and $c$ are numerical coefficients determined from the outcomes of numerical simulations of the sweep-up of orbiting debris.  In the canonical Moon-forming impact, the mass of escaping material is small, typically  $M_{esc} \leq 5$\% \citep{Kokubo2000}.  Pure $N$-body gravitational simulations that do not consider the role of gas in disk evolution yield $a=1.9$, $b=1.15$, and $c=1.9$ \citep{IdaCanupStewart}, which I refer to as the ``ICS97'' coefficients.  More realistic simulations that self-consistently model the evolution of the gas/vapor disk within the Roche limit along with the gravitational sweep-up of Roche-exterior material (see Section \ref{sec:disk-models}) yield $a=1.14$, $b=0.67$, $c=2.3$ (the ``SC2012'' coefficients, from \citet{SalmonCanup}), which yields a smaller resulting lunar mass for a given set of disk conditions.}

The mass fraction of iron in the final Moon is usually assumed to be equal to the mass fraction of iron in the disk \citep{Canup2004}.  One way of quantifying the mixing between the protoearth and impactor is to calculate $\delta F_T =(F_{D,tar}/F_{P,tar} - 1)$, where $F_{D,tar}$ is the mass fraction of target silicate in the disk, and $F_{P,tar}$ is the mass fraction of target silicate in the planet \citep{Canup2012}.  To calculate the value of $\delta F_{T}$ required to match the isotopic data, \citet{Canup2012} assumes that the magnitude of the difference in isotopic ratios between protoearth and impactor is similar to the difference between the isotopic ratios of the terrestrial mantle and Mars \citep{Zhang2012}.  {\Revision Given that the terrestrial planets, and by extension, the protoearth and impactor accreted from a common source of material \citep{RaymondMoon2015}, this assumption is a reasonable starting point for estimating isotopic differences between the two bodies that collided to form the Earth/Moon system.  However, it is an assumption that should be explored more fully in future work.}  In that framework, the most restrictive refractory isotope constraint comes from titanium \citep{Zhang2012}: $\delta F_T< 10$\%.  Another way of quantifying mixing is to simply calculate the mass fraction of disk material that originated from the projectile \citep{StewartCuk2012}, which should be less than 8\% to satisfy the titanium isotope constraints, {\Revision assuming that the impactor has an isotopic composition similar to Mars} \citep{Zhang2012}. 

\subsection{Toward a Scaling Relationship}
In addition to identifying candidate impacts that reproduce the physical, dynamical, and chemical constraints on the formation of the Moon, a secondary goal is to produce a scaling relationship between impact conditions and disk mass.  Relationships such as equation (\ref{eq:ICS97}) could then be used to predict the mass of the Moon.  Such a relationship would permit workers to predict the mass of Moon formed in a giant impact without having to perform a simulation.  

The amount of material launched into orbit in a planetary-scale collision depends on the impact geometry, velocity, and objects' compositions.  The disk mass, $M_D$, depends on the mass of the lens-shaped region representing the overlap between the target and impactor, $M_{interact}$ \citep{Canup2008, Leinhardt2012I}, 
\begin{equation}
\frac{M_D}{M_T} \sim C_{\gamma} \bigg(\frac{M_i - M_{interact}}{M_T}\bigg)^2, \label{eq:diskmass}
\end{equation}
with the factor $C_{\gamma} \sim 2.8 (0.1/\gamma)^{1.25}$ determined empirically based on the results of impact simulations \citep{Canup2008}.  A geometrical definition of $M_{interact}$ is shown in Figure \ref{fig:geometry}b.  \citet{Canup2008} shows that this relationship predicts $M_D$ to within a factor of 2 for impacts between {\Revision differentiated} 70\% dunite and 30\% iron targets for impact angles between 23$^{\circ}$ and 44$^{\circ}$, $1 \leq v_{imp}/v_{esc} \leq 1.4$, and for impact angles between 23$^{\circ}$ and 53$^{\circ}$ for $1 \leq v_{imp}/v_{esc} \leq 1.1$.

\subsection{Results \label{sec:GIs}}
\citet{CanupAsphaug} presented the first set of modern SPH simulations that demonstrated that a planetary collision could create a disk {\Revision with enough mass and the proper level of iron depletion} to make the Moon.  These conditions have come to be known as the ``canonical'' Moon-forming impact, and give (1) the proper masses for the Earth and Moon, (2) a system angular momentum consistent with the present value, and (3) a Moon with only 8\% iron by mass.  \citet{CanupAsphaug} showed that successful impacts had $\gamma \sim 0.1$ to 0.11, $v_{imp}=v_{esc}$, a total mass $M_T \sim 0.97$ to 1.02 Earth masses, and moderate impact angles.  These simulations were quickly followed up by \citet{Canup2004}, who used double the number of SPH particles (yielding higher numerical resolution), and an M-ANEOS equation of state for dunite, to show that successful collisions involve $\gamma\sim 0.11$ to 0.14, $v_{\infty} < 4$ km s$^{-1}$, and impact angles between 42$^{\circ}$ and 50$^{\circ}$.  A further numerical study showed that pre-impact rotation in the protoearth did not significantly affect the outcome of the canonical impact \citep{Canup2008}.  One example of a successful canonical Moon-forming impact, run119, simulated in CTH, is shown in Figure \ref{fig:ser119}.   Run119 involves a total mass $M_T=1.02M_E$, impactor-to-total mass ratio of $\gamma=0.13$, impact velocity $v_{imp}=v_{esc}$, an impact angle of 46$^{\circ}$, and that {\Revision both bodies have a differentiated interior composed of 70\% dunite and 30\% iron by mass}.

In the canonical impact scenarios, the overwhelming majority of material launched {\Revision into orbit} originates from the impactor.  The simulations of \citet{Canup2004} show that in all of the canonical scenarios, more than 70\% of the {\Revision initial disk material} originates from the impactor.  More head-on impacts can yield disks with less impactor material, but require larger impact velocities to loft enough material into orbit to accrete a Moon with the proper mass.

\section{{\Revision Scenarios Constrained by} Mass, Angular Momentum, Iron Fraction, and Chemistry   \label{sec:chemistry}}
\subsection{Isotopic Ratios}
At the same time that the canonical impact became solidified in the minds of dynamicists and modelers as the leading scenario for lunar origin, geochemical data began to cast doubt on its viability.  Writing in an abstract for the Annual Meteoritical Society Meeting in 2009, Jay Melosh declared the Giant Impact theory to be in ``crisis'' due to its inability to match the stable isotope data.  Melosh urged researchers to strive to resolve the crisis by re-visiting assumptions made in prior works, including the numerical methods used to simulate the impact itself, the assumption that the system angular momentum post-impact should be equal to the system angular momentum today, and the behavior of turbulent mixing in the disk post-impact \citep{Melosh2009}.  

{\Revision The similarities in isotopic ratios could be explained in three possible ways: (a) much of the Moon is actually terrestrial mantle material; (b) isotopic equilibration during the Giant Impact event; or (c) that the Earth and the impactor that created the Moon formed from isotopically similar material \citep{Wiechert2001}.}  All three scenarios have been explored recently using modern techniques.  In Sections \ref{sec:2012} and \ref{sec:equil} I describe hypotheses (a) and (b) in detail, because they are, at this point, the most well-developed.  

Hypothesis (c) has been explored by a single suite of $N$-body simulations of the late stages of accretion, which show that the Earth and impactor could originate from a common source of material \citep{RaymondMoon2015}.   The $N$-body simulations use ten times more particles to represent material in the terrestrial planet-forming region than any prior study.  Having many more particles increases eccentricity damping so that material accretes from a narrow ``feeding zone.'' Moreover, assembling each object from {\Revision more numerous} small objects increases the likelihood of compositional homogenization.  

However, this scenario does not provide a ready explanation for a recently discovered slight discrepancy between the tungsten isotope ratios of the Earth and Moon \citep{Touboul2015, Kruijer2015}.  The Moon has $\sim 27$ ppm excess $^{182}$W relative to the present bulk silicate Earth, which can be explained by accretion of a late veneer of chemically and isotopically distinct material \citep{Touboul2015}.  These works find \emph{no} difference in tungsten isotope ratios for pre-late-veneer Earth, despite the fact that the giant impact should change the isotopic ratios.  During the giant impact, tungsten isotopes from the impactor mantle would have been deposited in the Earth's mantle, and the impactor's core would have partially equilibrated with the terrestrial mantle as the impactor core sinks to merge with the terrestrial core \citep{Kruijer2015}.  Both of these processes can effect $\varepsilon^{182}$W.  Thus, isotopic ratios of the  impactor and protoearth could not have been \emph{identical} pre-impact; this issue has yet to be addressed {\Revision by $N$-body simulations.}

\subsection{High Angular Momentum Impact Scenarios \label{sec:2012}}
The isotopic data motivated a series of studies exploring alternate impact scenarios.  One possibility is that the Moon formed via a high-velocity, grazing ``hit-and-run'' impact in which a portion of the impactor leaves the final system, carrying away the excess angular momentum \citep{ReuferMoon2012, Meier2014}.  Multiple impacts is another possibility \citep{CitronMultipleImpacts, RufuMoon2015}.  These scenarios have difficulty creating moons of the proper {\Revision mass, angular momentum, and iron fraction.} 

A single impact event, but perhaps with different impact properties than the canonical model (velocity, protoearth-to-impactor mass ratio, impact angle, composition, or material properties) remains the most plausible explanation.  The two most successful alternative scenarios involve impacts that leave the Earth-Moon system with {\Revision an} angular momentum substantially higher than its present value.  The angular momentum constraint can be relaxed if the system can lose angular momentum after the Giant Impact, {\Revision during early tidal evolution \citep{KaulaYoder1976}.}  This situation is possible if the Moon is captured into the ``evection resonance'' {\Revision \citep{ToumaWisdom1998}}.  {\Revision The evection resonance is a dynamical state in which the period of precession of the lunar pericenter is similar to Earth's orbital period.}  When this commensurability occurs, the rotation rate of the Earth and the lunar semi-major axis decrease until the orbital frequency of the Moon near its pericenter matches the rotation rate of the Earth \citep{ToumaWisdom1998}.  The net result is that angular momentum from the Earth/Moon system is transferred to the Earth's orbit around the Sun.  This {\Revision may} allow for a variety of different formation scenarios to be viable, including fission {\Revision \citep{KaulaYoder1976}} which had been previously ruled out by angular momentum considerations.  

\begin{figure}[ht!]
\centerline{\includegraphics[width=80mm]{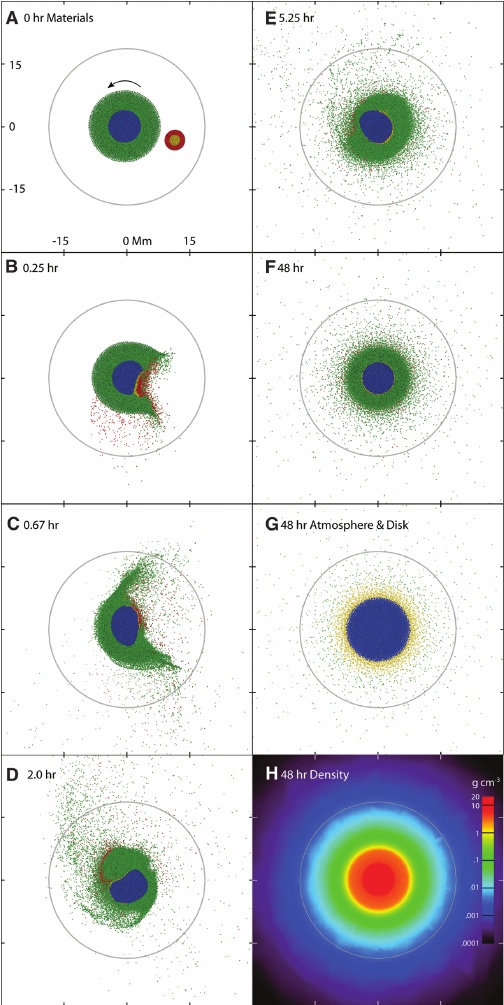}}
\caption{Snapshots of the first 48 hours of an SPH simulation of the formation of an impact-generated disk from a 0.05 Earth-mass impactor colliding with a protoearth of 1.05 Earth masses, spinning with a rotation period of 3 hours, at $v_{imp}=20$ km s$^{-1}$, from Figure 1 of \citet{StewartCuk2012}.  The resulting disk has about two lunar masses worth of material.  In panels A through G, colors  represent the provenance and composition: blue/yellow dots are the iron cores of the protoearth and impactor, and green/red dots are the protoearth and impactor mantles.  Panel G separates material into three categories: planet (blue), atmosphere (yellow), and disk (green).  In panel H, colors indicate densities, on a logarithmic scale.    \label{fig:csfigure1}}
\end{figure}

\citet{StewartCuk2012} used GADGET simulations to show that high-velocity impacts ($v_{imp} \sim 30$ km/s) of a small ($M_i\sim 0.026$ to 0.1$M_E$) impactor hitting a rapidly spinning Earth can yield a disk with enough material mixing between the bodies to account for the measured isotopic ratios.  
Figure \ref{fig:csfigure1} depicts {\Revision the results of a successful impact from their study: the first 48 hours of the collision of a 0.05 Earth-mass body onto a rapidly spinning Earth.}  The disk angular momenta in their successful cases ranges from $\sim 2$ to 3 $L_{EM}$, much higher than the canonical impact.  Their dynamical simulations show a high probability of capture into the evection resonance if the Earth is spinning with a rotation period less than 5 hours, and if the Moon has an initial semi-major axis $\sim 7 R_E$, which \citet{StewartCuk2012} report are met by their SPH simulations.  

\begin{figure}[ht!]
\centerline{\includegraphics[width=80mm]{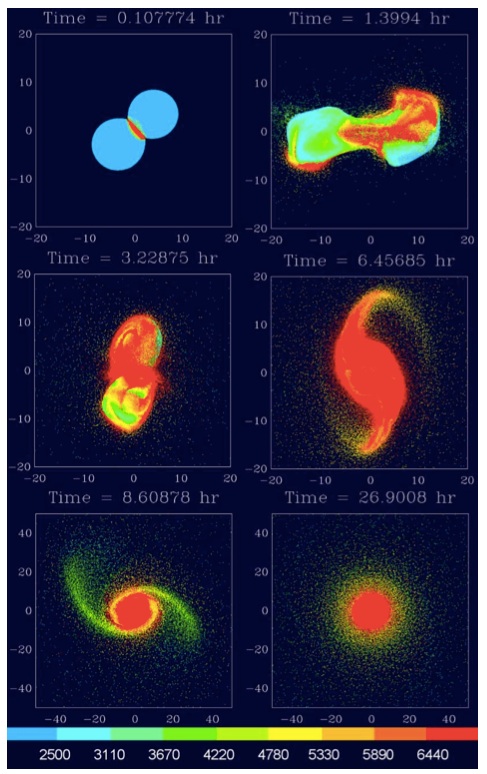}}
\caption{Snapshots of the first 30 hours of an SPH simulation of a ``half-Earth impact'' ($\gamma=0.45$, $v_{imp}/v_{esc}=1.1$, $v_{\infty}=4$ km/s, and impact angle$\sim 33^{\circ}$), reproduced from Figure 1 of \citet{Canup2012}.  Colors indicate temperatures in Kelvin.  These impact conditions lead to the formation of a 3 lunar-mass disk with angular momentum $L_D=0.47 L_{EM}$.  The protoearth and impactor have sufficiently mixed during the impact to explain similarities in the  titanium and chromium isotopic ratios.  The predicted moon mass, determined using equation (\ref{eq:ICS97}) with the ICS97 coefficients, is 1.64 lunar masses.  The impact leaves the Earth spinning with a 2 hour rotation period. \label{fig:run31}}
\end{figure}

\begin{figure}[ht!]
\centerline{\includegraphics[width=140mm]{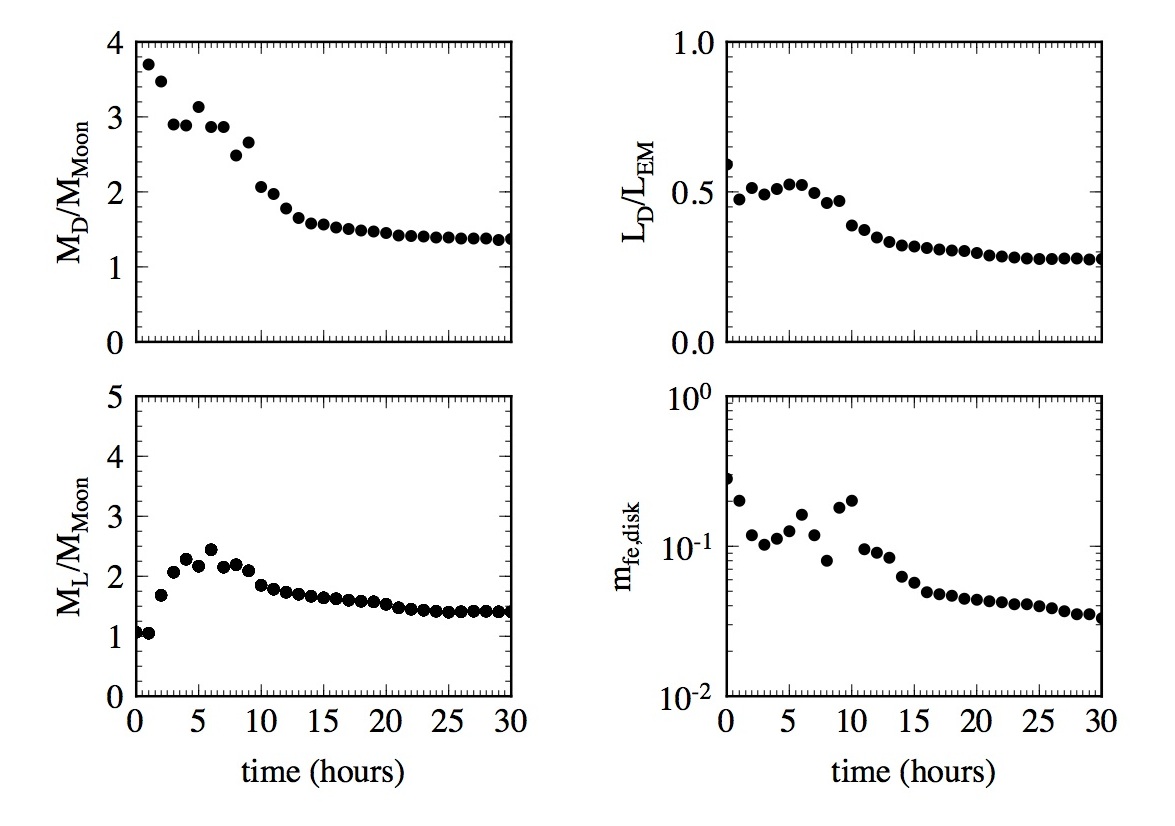}}
\caption{Evolution of the disk mass ($M_D$, top, left) scaled by the mass of the Moon ($M_{Moon}=7.35 \times 10^{25}$ grams),  disk angular momentum scaled by the present angular momentum of the Earth-Moon system ($L_{EM}=3.5 \times 10^{41}$ g cm$^2$ s$^{-1}$, top, right), the predicted lunar mass based on equation (\ref{eq:ICS97}) with ICS97 coefficients ($M_L$ bottom, left), and the mass fraction of iron in the disk ($m_{fe,disk}$, bottom, right), for the first 30 hours of the canonical Moon-forming impact simulation shown in Figure \ref{fig:ser119}.  After $t\sim 10$ hours, most of the orbiting clumps of material have either re-impacted the Earth or are in stable orbits.  The roughly linear decrease in the disk mass is due to spreading of the disk from the artificial viscosity in CTH \citep{m-moon}.
\label{fig:ser119-timelogs}}
\end{figure}

In a second paper, appearing back-to-back with \citet{StewartCuk2012}, \citet{Canup2012} shows that less-grazing impacts at $v_{imp} \sim 1.5 v_{esc}$, can yield, via a merging event, a much hotter, and more massive disk {\Revision composed of a mixture of material from both bodies, rather than primarily derived from one object, as in} the canonical impact.  Figure \ref{fig:run31} depicts a successful impact scenario from this work, with $M_T=1.04M_E$, $\gamma=0.45$, $v_{imp}/v_{esc}=1.1$, $v_{\infty}=4$ km/s, and impact angle$\sim 33^{\circ}$.  {\Revision Here, the two bodies collide multiple times, resulting in a merger, and lofting of silicate-rich mantle material off the merged body to make an orbiting disk. }  These impacts have come to be known as ``half-Earth'' impacts, because the impactor and target are close in size, each containing about half the mass of the Earth.  This work also relies on the evection resonance to shed excess angular momentum post-impact.  

These two studies seemed to match four of the five constraints on lunar origin, but subsequent work has cast doubt on the viability of the evection resonance as a means of changing the system's angular momentum post-impact.  The initial model of the dynamical evolution of the system into the evection resonance from \citet{StewartCuk2012} used several simplifying assumptions, each of which can lead to unphysical behavior.  These include a tidal torque that is always aligned with the spin angular momentum of the Earth rather than depending on the orbital elements of the Moon and a point-mass Moon, rather than a triaxial body with its own tidal/rotational bulges that exert a torque on the Earth \citep{Wisdom2015}.  With these assumptions relaxed, and considering tidal dissipation in the Moon with a simple constant-$Q$ model, \citet{Wisdom2015} find that the evection resonance, as envisioned by \citet{StewartCuk2012}, extracts \emph{too much} angular momentum from the system.  \citet{Wisdom2015} {\Revision identify} an alternate dynamical evolution pathway, {\Revision in which the} resonant argument of the evection resonance circulates, rather than oscillates.  This mechanism permits the angular momentum of the system to be decreased from the post-impact valued identified by the \citet{StewartCuk2012} simulations to its present value.  

\label{sec:disk-models}
More detailed modeling of the protolunar disk and the coupled thermal/orbital evolution of the Earth and Moon shows that the Moon may not be captured into the resonance in the first place \citep{SalmonCanup2014, Zahnle2015}.  Protolunar disk models for the hot, massive disks formed by the half-Earth SPH simulations show that the evection resonance may sweep past the Moon during the late stages of its accretion when the gas disk is dissipating, preventing the system from entering the resonance \citep{SalmonCanup2014}.  At this time, the Earth is still molten and essentially inviscid and so cannot undergo tidal heating; the Moon is only partially molten and would experience significant tidal dissipation, preventing the rise in eccentricity required for evection resonance capture \citep{Zahnle2015}.  

\subsection{The Protolunar Disk \label{sec:disk-structure}}
If the material that ultimately accretes onto the Moon spends a significant amount of time in physical contact with the silicate atmosphere of the Earth, this raises the possibility that isotopic ratios of lunar and terrestrial material could equilibrate \citep{PahlevanStevenson2007}.  However, modeling of the evolution of the disk as a purely particulate medium \citep{IdaCanupStewart} show that the material that ultimately accretes into the Moon originates from the outermost edge of the disk.  Thus, isotopic equilibration requires two conditions: (1) lateral transport of material within the disk, and (2) prolonged disk lifetimes {\Revision \citep{ThompsonStevenson, Ward2011, Charnoz2015}}.

\begin{figure}[h!]
\centerline{\includegraphics[width=100mm]{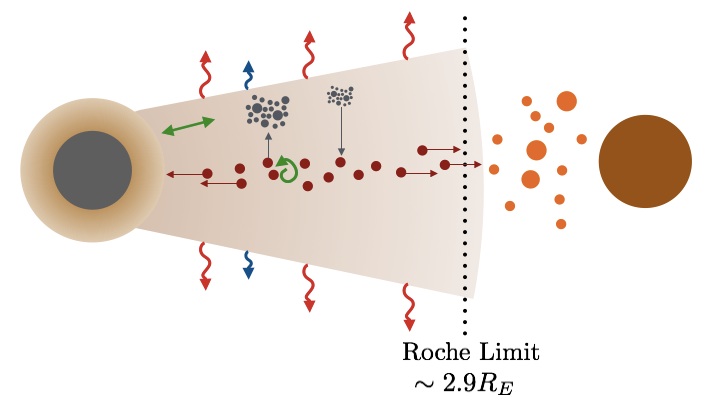}}
\caption{Edge-on schematic illustration of the modern view of processes at work in the impact-generated disk.  Inside the Roche limit (orbital distances less than 2.9 Earth radii), tidal forces prevent accretion, and the disk contains clumps of silicate melt (red) and vapor (brown cloud).  Vapor can condense, and melt clumps can be vaporized (gray arrows).  The disk cools from radiation (red squiggly arrows), and loses water (blue squiggly arrows).  The melt layer spreads diffusively, toward the protoearth, and beyond the Roche limit.  Turbulent mixing (green) may permit stable isotope ratios to equilibrate between the silicate atmosphere of the protoearth and the disk.  Beyond the Roche limit, the disk fragments into $\sim$ hundred kilometer-scale clumps (orange), which accrete onto the Moon (brown). \label{fig:disk}}
\end{figure}

Figure \ref{fig:disk} illustrates our modern understanding of the structure of the protolunar disk and its evolution for the $\sim10$ to 1000 years after the Giant Impact, while the Moon is accreting \citep{WardCameron1978, ThompsonStevenson, Ward2011}.  An excellent quantitative description of the protolunar disk may be found in \citet{Ward2011}.  Here, I describe the disk behavior in a qualitative sense, with the purpose of gaining a physical understanding what controls the lifetime of the disk, which in turn controls the length of time in which isotopes can equilibrate between lunar and terrestrial material, and the time scale for lunar accretion.    

After the impact, the debris quickly settles into a disk, located within a few Earth radii of the planet \citep{WardCameron1978}.  Inside the Roche limit, $\sim 3$ Earth radii, gravitational tidal forces from the Earth are too high to permit material to accrete.  Outside the Roche limit, material can clump up due to gravitational forces.  As it evolves and cools, the disk spreads inward toward the Earth, with some of the orbiting material accreting back onto the Earth, and outward, beyond the Roche limit \citep{Ward2011}.  As material from the disk moves beyond the Roche limit, it begins to fragment into clumps of initially molten material that cool (orange spheres in Figure \ref{fig:disk}) as they collide with each other, and collide with any large fragments of intact debris launched into orbit by the impact (brown).  The behavior of the disk inside the Roche limit is governed by radiative cooling, phase transitions, and turbulent convection \citep{ThompsonStevenson, Ward2011}, whereas the behavior of the Roche-exterior disk, where the gas density is small, is governed by gravitational forces.  

 The surface mass densities in the disk are quite high, as well, and can be $\sim 10^7$ g cm$^{-2}$ \citep{Ward2011}. A purely {\Revision particulate} disk with such a high surface mass density would be unstable and clump up, but tidal stresses would shear out the clumps as they form, generating an enormous effective viscosity for the disk \citep{WardCameron1978}.  A disk of this type spreads in a year or two, {\Revision and would result} in extremely rapid accretion for the Moon \citep{IdaCanupStewart, Kokubo2000}.  This would not permit any isotopic equilibration between lunar and terrestrial material, and {\Revision would predict} the formation of a completely molten Moon, which is inconsistent with limits on melting in the Moon implied by GRAIL data.  {\Revision However, such rapid disk evolution is unrealistic because the release of gravitational energy as heat would vaporize the disk material, preventing solids from forming.}

{\Revision The Roche-interior region of the protolunar disk is hot, with temperatures in the few thousands of Kelvin \citep{ThompsonStevenson, Canup2004, Ward2011}.} {\Revision A more physically realistic model} for the disk was proposed by \citet{ThompsonStevenson}, who suggested that the disk self-regulates due to a balance between {\Revision kinetic} energy dissipation and radiative cooling to exist in a metastable state.  In such a disk, the temperatures are $\sim 2000$ K, and the silicate material in the disk exists as a two-phase vapor/liquid state, akin to the ``frothy `head' on a beer'' \citep{ThompsonStevenson}.   

{\Revision The two-phase metastable state envisioned by \citet{ThompsonStevenson} can be maintained only if the disk has a very low mass of gas.  Subsequent work by \citet{Ward2011} proposes a structure in which drops of liquid settle to the midplane,} and the mass of gas is regulated by the equilibrium between radiative cooling and viscous dissipation {\Revision in the melt layer alone}.  This is the disk structure depicted in Figure \ref{fig:disk}. 

{\Revision In the \citet{Ward2011} scenario, the }Roche-interior disk consists of two reservoirs: a silicate vapor atmosphere, which is physically connected to the thick silicate vapor atmosphere of the young Earth \citep{Zahnle2015}, and a sheet of liquid silicate droplets at the midplane \citep{MachidaAbe2004, Ward2011}. The sheet of liquid silicate magma at the disk midplane spreads inward toward the Earth, and outward beyond the Roche limit, on a time scale $\sim$ 50 years (red arrows).  As the melt layer spreads, the rate of viscous dissipation in the disk decreases, causing the disk to cool, and condense, producing more liquid, which resupplies the melt sheet.  This process continues until all of the vapor has condensed, {\Revision $\sim 250$ years.}

{\Revision The} rate of radiative cooling controls the rate at which the disk spreads beyond the Roche limit, and the time scale for lunar accretion. Thus, the Moon can form in hundreds of years, rather than a few years, as predicted by models of a purely {\Revision particulate} disk {\Revision \citep{WardCameron1978}}.  This leaves more time for isotopic equilibration between the disk and the terrestrial silicate atmosphere, and for the disk material to cool before accreting onto the Moon. 

{\Revision Recently, \citet{Charnoz2015} have simulated the evolution of the protolunar disk using a one-dimensional time-dependent model including viscous heating, radiative cooling, phase transitions, and gravitational instability.  Their results show an even more prolonged disk evolution than implied by recent works: the disk solidifies in $10^3$ to $10^5$ years, permitting more time for isotopic equilibration and cooling of disk material.  They find that a balance between radiation and dissipation is never achieved, and that radiative cooling is always more effective than viscous dissipation.}

\subsection{Isotopic Equilibration {\Revision With} the Disk \label{sec:equil}}
\citet{PahlevanStevenson2007} show that isotopic equilibration can be achieved due to mixing of the disk via turbulent convective motions driven by the disk's rapid rate of cooling.  They consider mixing across the disk to be described by Fick's law, which describes the diffusive mixing of materials of two compositions.  The diffusion, in this case, is accomplished by turbulent eddies, whose characteristic diffusivity is described $D=\alpha c_s H$, where $c_s$ the sound speed of the disk, $H$ is its scale height, and $\alpha$ describes the vigor of fluid motions.  Solutions of the diffusion equation show that sufficient equilibration can occur if the disk lasts for 10 to 1000 years, and $\alpha \sim 10^{-3}$ to 10$^{-4}$.  Values of $\alpha$ of this magnitude are possible if the length scale of the turbulent eddies is about one tenth of the scale height of the disk.  It is important to note that although \citet{PahlevanStevenson2007} focus on equilibration of oxygen isotopes, nothing in their method depends on the specific isotope.  \citet{Pahlevan2011} point out that their methods can be applied generally to diffusion of any species across the disk, including bulk chemical species. 

For isotopes of very refractory species such as titanium to be equilibrated, the disk would need to survive for longer than \citep{Zhang2012},
\begin{equation}
\tau_{ex} = \frac{C \sigma}{P_v} \sqrt{\frac{2 \pi R_GT}{m}},
\end{equation}
where $\tau_{ex}$ is the timescale for exchange of material between the melt sheet and the disk vapor, $P_v$ is the vapor pressure of the element, $C$ is the element concentration in the disk, $R_G$ is the gas constant, $\sigma$ is the surface mass density of the disk, $T$ is the temperature, $m$ is the {\Revision molar mass for a given atom}.  For tungsten, $\tau_{ex} \sim 1$ year, and for titanium, $\tau_{ex}\sim 30$ years \citep{Zhang2012}.  {\Revision Alternatively, fluid motions (e.g. convective overshoot) can directly carry the liquid into the vapor atmosphere where refractory elements have an opportunity to equilibrate via the exchange of droplets coupled to the turbulent vapor \citep{Pahlevan2011}.}

{\Revision  One requirement of this scenario is that the turbulent motions responsible for mixing of the disk must overwhelm the processes responsible for the transport of angular momentum \citep{Melosh2009}.}  
It is important to note that in the \citet{ThompsonStevenson} and \citet{Ward2011} disk models, the physical mechanism responsible for spreading in the protolunar disk {\Revision (i.e., gravitational instability)} is \emph{not} the {\Revision main mechanism} advocated for spreading in astrophysical disks (i.e., {\Revision magneto-rotational instability}).  The lifetime of the disk is controlled by the rate at which {\Revision it radiates the gravitational energy liberated by viscous spreading \citep{ThompsonStevenson, Ward2011}.}  Thus, the physical process controlling the disk spreading is not the same process controlling the convective mixing.  To understand the potential for isotopic equilibration, {\Revision it is} crucially important that the physical {\Revision mechanisms for spreading of the protolunar disk, and the physical mechanisms responsible for the mixing of trace species, be clarified.}


\subsection{Volatile Depletion}
The \citet{SalmonCanup} simulations show that the Moon is assembled in three phases: (1) in the first few years after the Giant Impact, material outside the Roche limit accretes into a lunar ``parent body'' about half the mass of the Moon; (2) growth stalls for a few tens to a hundred years as the inner disk spreads outward due to its viscosity; (3) as the disk spreads beyond the Roche limit, it spawns small bodies (``moonlets'') that accrete onto the parent body in a hundred to a few hundred years \citep{SalmonCanup}.  {\Revision Time-dependent models of disk evolution show that even longer lifetimes of $10^4$ to $10^5$ years may be possible \citep{Charnoz2015}.}

\citet{Canup2015} have recently modeled the loss of volatiles in an anhydrous disk, within the framework of the \citet{Ward2011} protolunar disk model.  The simulations show that Roche-interior {\Revision melt} would be depleted in volatiles, owing to the high temperatures in this region of the disk.  As the disk spreads and cools, more volatiles can condense.  The Moon would then be composed of parent body material originating from outside the Roche limit, covered by a layer of volatile-depleted material originating from the initial spreading of the Roche-interior disk.  During the late stages of disk evolution, more volatile-rich disk material would eventually accrete onto the Earth, {\Revision causing} the difference in volatile abundances between the two bodies. {\Revision \citet{Charnoz2015} show that if the vapor layer of the protolunar disk is viscous, much of the volatile-rich material falls onto the Earth's surface in the first ten years of disk evolution.}

Although this provides a {\Revision possible} explanation for the elemental abundances of volatile species, it does not yet provide an explanation for the high lunar water content.  {\Revision \citet{PahlevanKaratoFegley} have recently studied the partitioning of hydrogen between vapor and melt in the disk, and found that enough H can be dissolved in the vapor to explain the H abundance in the lunar interior.}  Another possibility is that some of the water in the lunar interior was delivered by cometary impacts to the lunar magma ocean \citep{Barnes2016}.  The extent of cometary and asteroidal bombardment required to supply the $\sim$ tens to hundreds of parts per million of water implied by lunar samples is consistent with dynamical models of the bombardment rates onto the young Moon \citep{Barnes2016}.

\section{The Moon's Initial Thermal State \label{sec:assembly}}
\begin{figure}[h!]
\centerline{\includegraphics[width=100mm]{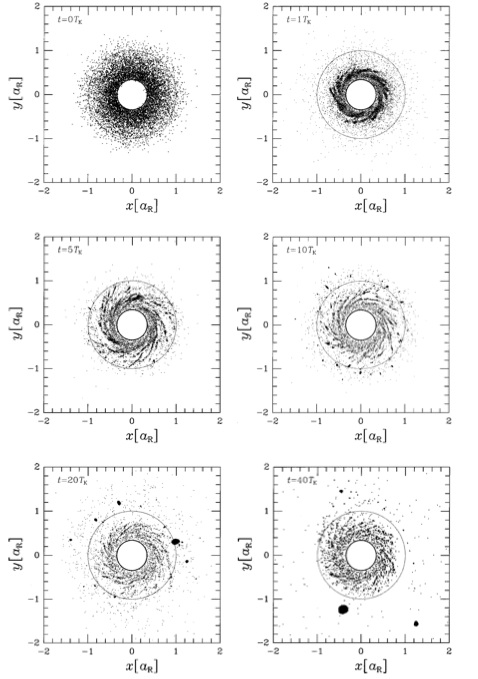}}
\caption{Snapshots of the first 12 hours of gravitational $N$-body simulations of the formation of the Moon from solid particles of impact-generated debris, reproduced from Figure 4 of \citet{Kokubo2000}.  Time is scaled by $T_K$, the Keplerian orbital period at the Roche radius ($T_K \sim 7$ hours).  The disk of material is projected into the $x-y$ plane, and the solid circle represents the Earth, the dotted circle is the Roche limit.  The disk initially has 4 lunar masses of material, represented by 10,000 solid, equal-mass particles. \label{fig:kokubo}}
\end{figure}

The accretion of the impact-generated debris into the Moon has been studied using numerical simulations with increasing complexity and realism since the mid 1990's.  One difficulty of simulating the sweep-up of debris post-impact is properly modeling the behavior of the disk inside the Roche limit.  One simplifying approach is to assume that the Roche-interior disk has solidified completely, and treat all of the mass as solid particles, subject only to gravitational forces (e.g., \citet{IdaCanupStewart, Kokubo2000}).  However, even this approximation requires a careful treatment of collisions between particles.  \citet{CanupThesis} points out that bodies of vastly differing size can accrete a little bit inside the classical Roche limit, even though like-sized bodies can not.  Additionally, particles that collide and rebound with velocities less than their mutual escape velocity can still be subject to accretion \citep{Canup1996}.  Using these ideas in a gravitational $N$-body simulation, \citet{IdaCanupStewart} showed that the mass of Moon created depends on the mass of the orbiting disk with a simple algebraic relationship (equation \ref{eq:ICS97}). Higher-resolution $N$-body simulations by \citet{Kokubo2000} (see Figure \ref{fig:kokubo}) show that, in the absence of gas, the impact-generated debris would form a spiral-arm structure, with moonlets forming at the tips of the spirals, finally accreting into a single Moon on a $\sim 1$ month time scale.  

\begin{figure}[h!]
\centerline{\includegraphics[width=80mm]{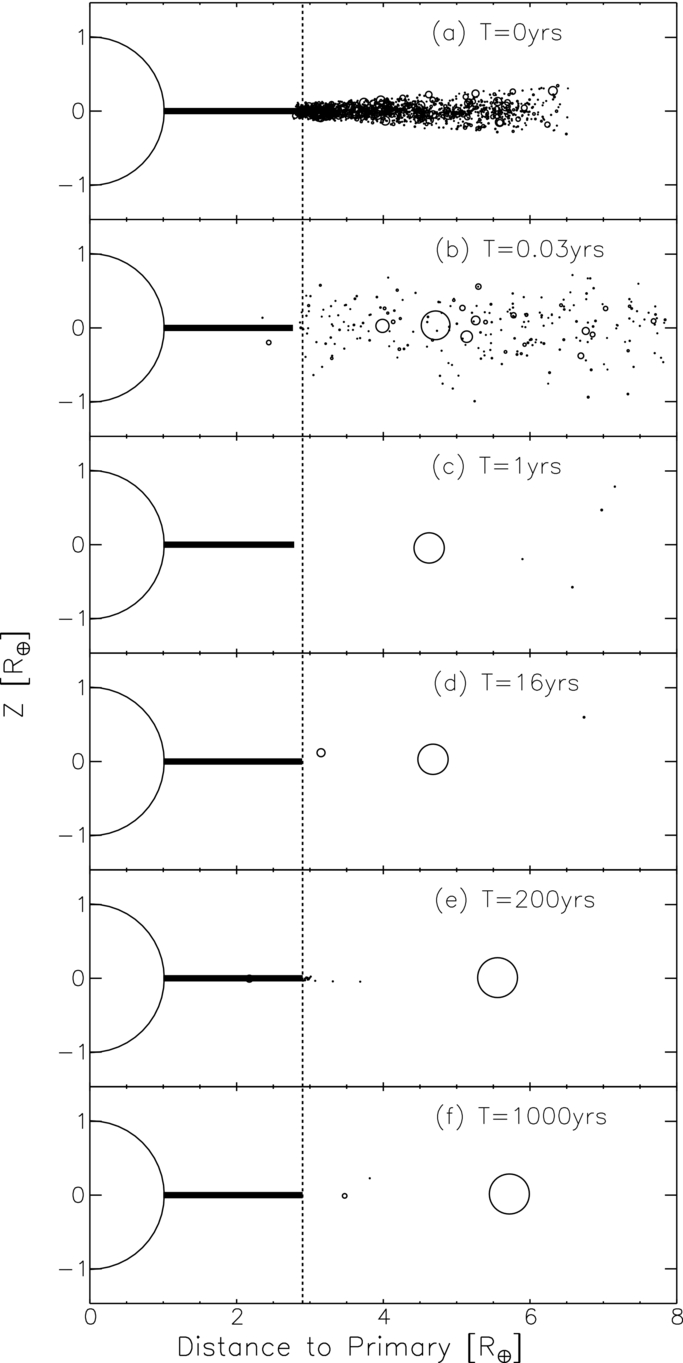}}
\caption{The protolunar disk, viewed edge-on, from the simulations of \citet{SalmonCanup} (reproduced from their Figure 2).  The plots show the $R-z$ plane of the disk and evolution of the system from the initial state, to $t=1000$ years, with a single large Moon.  The simulation represents the Roche-interior disk (distance to primary less than 2.9 Earth radii, vertical line) with an analytic description of the disk consistent with \citet{ThompsonStevenson} and \citet{Ward2011}, and the Roche-exterior disk with an $N$-body gravitational model.  \label{fig:SC2012}}
\end{figure}

However, in the modern vision of the protolunar disk, the time scale for lunar formation should be controlled by the rate of supply of material from the Roche-interior disk, which is of order tens to hundreds of years, not by the orbital period of the debris.  A hybrid approach has been recently developed to simulate both the Roche-interior and Roche-exterior portions of the disk self-consistently \citep{SalmonCanup}.  The Roche-interior disk is described using the analytic model of \citet{Ward2011}.  As material from the Roche-interior disk spreads beyond the Roche limit, new particles are added to the $N$-body gravitational simulation.  Roche-interior material that spreads toward the protoearth is added to the Earth and removed from the disk.  Figure \ref{fig:SC2012} illustrates the time-evolution of the disk and accretion of a single Moon in 1000 years.  The stall in growth rate, and the formation of small moonlets, provides hope that the Moon can cool sufficiently during its accretion to remain {\Revision only} partially molten, potentially allowing minimal melting during formation, consistent with the thickness of the Moon's primary crust.



\subsection{Background}
A body as large as the Moon will release enough gravitational potential energy during its formation to melt completely \citep{StevensonSatsBook}, unless it forms slowly from small objects that deposit their energy close to the surface where it can be lost, e.g., by radiation \citep{PS00, cr}.  The gravitational potential energy per unit mass liberated by the formation of the Moon, $E_g \sim \frac{3}{5} GM_{L}/R_{L} \sim 2L$, where $M_{L}$ and $R_{L}$ are the lunar mass and radius, and $L\sim 4 \times 10^5$ J/kg is the latent heat of silicate \citep{PS00}.  At face value, this suggests the Moon should melt completely during formation, regardless of its accretion conditions.  

To determine the temperature inside a growing planetary body, it is necessary to know how impact kinetic energy is deposited as heat, and how much heat is retained versus lost to cooling processes.  To date, all models of the thermal evolution of a growing planet use a similar approach: a one-dimensional energy balance that assumes a poorly defined and poorly constrained fraction, $h$, of the accretional energy is deposited as heat in the subsurface of the planet \citep{Kaula1979, RansfordKaula, Squyres88, Senshu, MerkPrialnik2006}.  Unfortunately, under most conditions, accretional temperature profiles are extremely sensitive to $h$ \citep{StevensonSatsBook, SchubertSatsBook}, and so the effect of changing $h$ to account for its unknown value significantly outweighs the effect of changing the accretion time scale or temperature of accreting bodies (see, e.g., \citet{Squyres88} for discussion).  With such a model, it is not possible to relate limits on early melting to the conditions of a planet's accretion.

Constraining $h$ is challenging, even with modern numerical techniques.  The value of $h$ likely depends on numerous properties of the growing planet, including the size, composition, and physical state of the impactors and the planet; the time scale for planetary growth; and the dynamical environment from which the accreting material is drawn (see e.g., \citet{MeloshOriginEarth} for discussion).  Additionally, the amount of heat from each impact {\Revision that} gets deposited in the growing planet likely {\Revision depends} on the strength of the surface of the planet \citep{GH, OKeefeAhrens1977}, which controls how much energy is used to heat versus break/crush the target.  Even CTH with its state-of-the-art material models, is not very well-suited to obtaining estimates of $h$ at the accuracy required for modeling accretional impacts because it conserves momentum, not energy \citep{CrawfordPers}.  Furthermore, at low velocities like those associated with lunar accretion, a few kilometers per second, the shock launched by an impact is weak, and comes after an elastic pre-cursor wave that compresses the target \citep{Melosh1989}; these processes are not well-represented in hydrocodes currently used to study planetary cratering.  


However, in considering the thermal state of a Moon accreted from a two-phase disk, we can learn from techniques used to simulate the formation of Jupiter's large, apparently undifferentiated ice/rock satellite, Callisto.  Callisto is another moon that seems to have experienced only limited melting during its formation and early history.  Measurements of its gravity field show that its polar moment of inertia coefficient is anomalously large \citep{JupBookInteriors}.  The simplest explanation for this is that part of Callisto still contains an intimate mixture of ice and rock \citep{JupBookInteriors}, indicating that it did not melt during its accretion, early history, or long-term thermal evolution \citep{McKinnonLPSC2006, cr}, despite the fact that the potential energy released during its accretion is six times the amount of energy required to melt its ice \citep{cr}.  

The work of \citet{cr} on the formation of an unmelted Callisto motivates a different approach that does not rely on $h$.  Rather than accurately determining the accretional temperature profile, this approach constrains the \emph{coldest} possible state for the growing Moon for a given set of accretion conditions, by assuming that all of the impact kinetic energy is deposited as heat near the Moon's surface, in a layer thin enough to cool between successive overlapping impacts.  Using this approach, I show that in the context of the modern two-phase disk models of \citet{Ward2011}, and the hybrid disk-$N$-body models of \citet{SalmonCanup}, the Moon may be able to avoid widespread melting, yielding a crustal thickness consistent with spacecraft data.  I construct analytical models of the thermal evolution of the growing Moon and the debris that accretes onto the Moon during the three phases of lunar assembly: formation of the parent body, cooling during the pause in accretion due to the spreading of the Roche-interior disk, and accretion of the Moon's outer layers from small objects spawned off the Roche-interior disk.  Each of the models assumes that cooling is maximally efficient, yielding the coldest possible thermal state for the Moon.

\subsection{Accretion and Cooling of the Lunar Parent Body}
As demonstrated by Figure \ref{fig:SC2012}, during the first phase of lunar accretion, a ``parent body'' composed of Roche-exterior debris forms.  The parent body is about  $40$\% the mass of the Moon, and has a radius $\sim$75\% of the lunar radius.  The object accretes in a few {\Revision months}, too quickly to permit any appreciable cooling.  The energy liberated during the assembly of the parent body is comparable to the latent heat of silicate, $E_g \sim L$, indicating it will be completely molten \citep{SalmonCanup}. 


After the parent body is assembled, it does not suffer any accretional impacts for a period of tens to hundreds of years while the Roche-interior disk is spreading \citep{SalmonCanup}.  During this time, the moonlet is essentially a floating magma ocean that cools due to vigorous convection in its interior.  The parent body will quickly develop a solid crust, which can limit the efficiency of convective heat transport.  If the crust remains intact, the surface heat flow from the magma ocean convection can be described using a relationship similar to the heat flow from stagnant lid convection \citep{ReeseSolomatov2006}.  However, if the lid periodically founders, the vigorously convecting magma/crystal mixture radiates its heat, giving rise to a high heat flow.  The heat flow in such a system is governed by the Rayleigh number, which describes the vigor of convection, and the Prandtl number, $Pr=\eta_l/\rho_l \kappa$, which describes the ratio of momentum to thermal diffusivity in a fluid.  Table \ref{table:magma} summarizes the thermal and physical properties of silicate magma, from \citet{ReeseSolomatov2006}.  Here, $\eta_l$ is the viscosity of the melt, $\rho_l$ is its density, and $\kappa$ is its thermal diffusivity, which yields $Pr \sim 30$.  

The heat flux for turbulent convection at high Prandtl number \citep{ReeseSolomatov2006},
\begin{equation}
F_{conv}=0.089 k (T-T_s)^{4/3} \bigg(\frac{\rho_l g \alpha}{\eta_l \kappa}\bigg)^{1/3},
\end{equation}
where $k$ is the thermal conductivity, $T$ is temperature of the magma, $T_s$ is the surface temperature, $g=1.17$ m s$^{-2}$ is the surface gravity on the parent body, $\alpha$ is the coefficient of thermal expansion.  The convective heat flux must be equal to the radiative heat flux from the surface of the magma ocean, which allows us to also estimate $T_s$: $F_{conv}=\sigma_{SB} T_s^4$, where $\sigma_{SB}$ is the Stefan-Boltzmann constant.  For a magma temperature $T=2000$ K, $T_s=1350$ K, and $F_{conv} \approx 2 \times 10^5$ W m$^{-2}$.  If the crust of the parent body remains intact, convective heat flow is a factor of $\sim 2$ to 3 lower \citep{ReeseSolomatov2006}.

The parent body will convect vigorously until it reaches the rheologically critical melt fraction, the fraction of melt at which the viscosity of the magma/crystal mixture becomes dominated by the solid crystals rather than liquid melt \citep{RennerEPSL}.  The time scale for the parent body to cool \citep{ReeseSolomatov2006},
\begin{equation}
\tau_{crys} \approx \frac{[L(1-\phi_{cr}) + C_{p,l} \Delta T_{cr}] \phi_l V}{4 \pi R_p^2 F_{conv}},
\end{equation}
where $L$ is the latent heat of silicate, $\phi_{cr}$ is the rheologically critical melt fraction, $\Delta T_{cr}$ is the temperature difference associated with cooling to $\phi=\phi_{cr}$, $V=(4/3) \pi R_p^3$ is the volume of the parent body, $R_p$ is its radius, and $C_p$ is the specific heat.  Parameter values are summarized in Table \ref{table:magma}.  The time scale for the parent body to cool to $\phi>\phi_{cr}$ is 100 years, indicating that the parent body is {\Revision partly} solidified by the time the Roche-interior disk spreads beyond the Roche limit and accretion continues.

\subsection{From Parent Body to Fully Assembled Moon}
In the final phase of lunar accretion, the parent body grows due to collisions with small moonlets, spawned at the edge of the Roche-interior disk \citep{SalmonCanup}.  The sizes of the objects are, plausibly, $10^{-9}$ to $3 \times 10^{-3}$ lunar masses, or roughly to 1 to 250 km in radius \citep{SalmonCanup}.  {\Revision By the time the disk has begun to spread beyond the Roche limit, $\sim 100$ to $10^4$ years \citep{ThompsonStevenson, SalmonCanup, Charnoz2015}, disk temperatures have cooled to $\sim 500$ to 1000 K \citep{Charnoz2015}.  Thus, objects accreting onto the parent body may have temperatures of this magnitude \citep{PS00}.}

If all accretional heat is lost by radiation, the satellite {\Revision temperature} $T$ as a function of radius $r$ is governed by \citep{cr},
\begin{equation}
\bar{\rho} C_{p,s} (T-T_i) \dot{r} =\frac{1}{2} \frac{\dot{M} v_i^2}{4 \pi r^2} - \sigma_{SB} T^4, \label{eq:simple_model}
\end{equation}
where $\bar{\rho}=3.32$ g/cc is the lunar density, $C_{p,s}=700$ J/kg-K is the specific heat of the solid Moon/moonlets, $T_i$ is the moonlet temperature, $T$ is the temperature at radius $r$, $\dot{r}=dr/dt $ is the rate of growth of the Moon's radius, $\dot{M}\approx M_{L}/\tau_{acc}$ is the mass accretion rate, and $v_i$ is the impact velocity of moonlets.  A Moon with a magma ocean (crystal fraction $\phi>$ 50\%) 300 km thick (to match GRAIL estimates; \citet{Andrews-Hanna2013}) would have $T \sim 1600$ K when $r=1438$ km, assuming that melt fraction scales linearly with temperature \citep{McKenzieBickle} between the liquidus and solidus for anhydrous peridotite ($T_{liq}=1921$ K and $T_{sol}=1334$ K; \citet{ZhangHerzberg}).  

Equation \ref{eq:simple_model} can be solved for $\tau_{acc}$ required to achieve these temperature and radius values by substituting $\dot{r}=r/(3\tau_{acc})$, and assuming the moonlet impact velocity $v_i^2 \sim v_{esc}^2 = 2 GM/r$ (cf. \citet{cr}),
\begin{equation}
\tau_{acc} = \frac{\bar{\rho}r[\frac{4}{3} \pi G \bar{\rho} r^2 - C_{p,s} (T-T_i)]}{3 \sigma_{SB} T^4}.
\end{equation}
This gives $\tau_{acc} \sim 200$ yr, similar to the time scale of the third phase of accretion estimated by \citep{SalmonCanup}, suggesting that a partially melted Moon could plausibly form from an impact-generated disk.

\section{Future Work \label{sec:future}}
The origin of the Moon has been, and continues to be, one of the richest and most challenging problems in Solar System geophysics.  With no comparable laboratory-scale experiments or field observations of planet-scale collisions and accretion possible, the community relies on numerical simulations and analytic theory to explain the observed properties of the system.  Lunar samples and meteorites provide small windows into the history of lunar material, but reconciling their compositions and isotopic ratios with the results of global-scale geodynamical and orbital dynamical modeling has been a challenge.  

However, in recent years, great strides have been made to construct a theory for lunar origin that is consistent with all of the geochemical and geophysical data available to us, not just the mass, angular momentum, and bulk lunar iron fraction.  

The most plausible explanation, at present, is that the Moon was formed in a Giant Impact between two large planetary embryos during the late stages of terrestrial planet formation in our Solar System.  The Moon and the impactor may have formed from isotopically similar material \citep{RaymondMoon2015}, but this does not explain the slight excess of $^{182}$W observed in the Moon \citep{Touboul2015, Kruijer2015}.  The impact launched a disk of hot, silicate-rich material into orbit around the young Earth.  The disk is another possible site of material exchange between lunar and terrestrial material.

The canonical Giant Impact can yield the proper mass, angular momentum, and bulk chemical composition for the Earth and Moon.  However, {\Revision the canonical impact alone cannot} explain the isotopic similarities between the Moon and the terrestrial mantle {\Revision because the impact-generated disk is formed primarily from impactor material.}  Recently proposed scenarios that yield more terrestrial-rich disk material leave the system with too much angular momentum.  Thus, they ignore the angular momentum constraint for the sake of getting the appropriate composition.  The detailed tidal evolution of the system post-impact remains an area of active research, with present studies suggesting that \citet{StewartCuk2012} {\Revision overestimate both the likelihood that the Moon could be captured into the evection resonance \citep{SalmonCanup2014}, as well as the efficiency of angular momentum transfer in the resonance \citep{Zahnle2015}.}  More sophisticated simulations of the system's evolution post-impact are required to fully assess the extent to which the angular momentum of the system can be changed post-impact.  

However, {\Revision fluid dynamical} processes occurring after the impact itself may have modified the chemical composition of the Moon to match observations.  Physical mixing and transport occurring during the impact and in the disk \citep{Ward2011} {\Revision can give rise} to the disparate chemical composition of the Earth and Moon \citep{Pahlevan2011, Canup2015}.  Similarities in the stable isotope ratios of both volatile and refractory species can be explained by equilibration in a long-lived protolunar disk \citep{Zhang2012}.  Volatile depletion in the Moon can be explained by the accretion of material spawned off the edge of a hot Roche-interior disk \citep{Canup2015}.  {\Revision Dissolution of water into disk melt may be able to account for the relatively high water content in lunar samples \citep{PahlevanKaratoFegley}.  Alternatively, water} delivery to the Moon by comets and asteroids may be able to account for the unexpectedly high water content in lunar volcanic glasses \citep{Barnes2016}. 

In spite of this tremendous progress in recent decades, some open questions remain:

\begin{itemize}
\item \emph{Is there a {\Revision self-consistent} single-impact {\Revision and disk evolution} scenario that is consistent with the masses, angular momentum, lunar iron fraction, and isotope/volatile compositions of the Earth and Moon?}  Answering this question may require significant computational power, improvements in the basic methods of solution for hydrocodes (so that shocks and disk evolution can be more accurately calculated), improved equations of state and material models, and a broader exploration of the parameter space of impact conditions.  In the hopes of encouraging more research groups to tackle this problem, I have provided, in the appendices of this paper, information to benchmark Giant Impact simulations against prior results.

\item \emph{Do the chemical compositions and isotopic ratios preserved in lunar samples and meteorites reflect the chemistry of the bulk Moon?}  The vast majority of lunar samples come from the upper $\sim 500$ km of the Moon \citep{TaylorTaylorTaylor}. The \emph{Apollo} samples were collected from a limited geographic region of the Moon, which may not reflect the bulk composition of the entire body (e.g., \citet{Haskin2003}).  Lunar meteorites can help alleviate this bias, but there are not many meteorites to study.  Samples of lunar mantle material from the South Pole-Aitken Basin would provide valuable insight into the composition of the deeper lunar interior \citep{TaylorTaylorTaylor}.  

\item \emph{What are the processes that control the spreading rate in the protolunar disk?}  {\Revision Does the disk experience an episode of mixing where the rate of diffusion of tracers overwhelms the transport of angular momentum?}

\item \emph{How does tidal dissipation in the Moon and Earth affect their very earliest phases of orbital evolution?}  Tidal dissipation in solid bodies is a notoriously complicated process, and not well-understood.  It is known in the terrestrial community that the tidal quality factor of a solid planet depends on frequency, as well as the physical state of the planet's interior.  Yet, many of the tidal models to date assume very simplistic interior states for the Earth and Moon.  Given recent advances in simulating tidal evolution of icy satellites {\Revision (e.g., \citet{Behounkova2015})}, and the recent work on {\Revision lunar accretion,} the tidal evolution of the Earth/Moon system is overdue for an update.

\item \emph{Is {\Revision the} initial thermal state of the Moon consistent with its formation from an impact-generated disk?}  This oft-overlooked constraint on the origin of the Moon can, and should, be addressed in the context of modern impact and planet formation theory.  Here, I have presented a mere plausibility argument, showing that the \emph{coldest} possible interior state for the Moon is consistent with limits on early melting based on its crustal thickness.  More sophisticated modeling in three dimensions is required to study this process further \citep{MeloshOriginEarth}.
\end{itemize}

\section*{Acknowledgments}
I am grateful to R. M. Canup, H. J. Melosh, D. J. Stevenson, and W. R. Ward for introducing me to the problem of lunar origin and for countless helpful discussions on this topic.  I also thank  M. Bruck, P. Schultz, and A. Stickle for sharing their thoughts on the potential role of strength in giant impact simulations. Reviews from Kaveh Pahlevan, an anonymous reviewer, and editor Steven Hauck significantly improved the clarity of the manuscript.  This work is supported by NASA Emerging Worlds grant NNX16AI29G.  Information about how to reproduce impact simulations shown in Figure 2 is listed in Appendices A and B.  Source data for Figure 5 is reproduced in Appendix B.  CTH input files used to create the simulation in Figure 2 are available upon request from the author.

\appendix
\section{ANEOS Equations of State  \label{sec:aneos-appendix}}
Here, I summarize the coefficients for the ANEOS equation of state for iron and dunite used by \citet{m-moon} to simulate the Moon-forming impact in SPH and CTH.  Table \ref{table:aneos-iron} lists the values of input parameters for ANEOS for iron.  These coefficients are slightly modified from the ``library values,'' (e.g., reported in \citet{ThompsonLauson1972}).  

Tables \ref{table:aneos-dunite} and \ref{table:aneos-dunite2} summarize the coefficients for the ANEOS equation of state for dunite, with molecular vapor.  This specific set of coefficients is designed to match those used by \citet{Canup2004}, for the purposes of cross-comparing CTH and SPH numerical results.  The coefficients are listed in the order in which they were input in CTH 9.1 \citep{m-moon}.  Researchers using other version of ANEOS or newer versions of CTH should take care to ensure that the parameters are entered in the proper order.  These have been previously reported in Appendix  A of \citet{m-moon}, but I include them here, also, for completeness, along with a slightly expanded discussion of the meaning of each parameter, and the differences between M-ANEOS and ANEOS.  

The classic ANEOS equation of state treats vapor as a mixture of atoms, which means that very high energies are required to create vapor.  As a result, the original ``atomic'' ANEOS may significantly underestimate vapor production in hypervelocity impacts \citep{MeloshMANEOS}.  \citet{MeloshMANEOS} modified ANEOS to treat molecular gasses, and presents a set of ANEOS coefficients for SiO$_2$.  \citet{m-moon} used his modified form of ANEOS (which is now distributed as part of CTH), his SiO$_2$ coefficients, and his prior ANEOS coefficients for dunite \citep{BenzMoonPaper3} to create a molecular equation of state for dunite.  

The overwhelming majority of these parameters originate from the \citet{BenzMoonPaper3} equation of state.  Key changes to the equation of state to treat molecules arise in the ``cold'' term of the Helmholtz free energy, which describes the behavior of the material at densities much larger than the reference density at zero temperature, and also at extremely low densities.  In the atomic EOS, the parameter $E_{sep}$ describes the energy required to separate the atoms from each other.  In the molecular version, this parameter is used to describe the energy to vaporize the material, $E_{vap}=1.3\times10^{11}$ erg/g. 

The molecular ANEOS also includes a modification to the inter-atomic potential for cold material (compressed at zero temperature, or very low densities).  Rather than using the Morse or Lennard-Jones potentials available in ANEOS, the updated version uses what Melosh refers to as ``Mie-type'' potential (of which Lennard-Jones is a specific case), {\Revision where the pressure} (cf., equation 4 in \citet{MeloshMANEOS}),
\begin{equation}
P_{cold}=C(\eta^m - \eta^a) \textrm{ for } 0 < \eta < 1,
\end{equation}
where $m>a$ to assure $P_{cold}$ is tensional, $\eta=\rho/\rho_o$ is the compression, $\rho$ is the material density, and $\rho_o$ is the reference density.  The constants $C$ and $m$ are determined from the vaporization energy (related to the integral of $P_{cold}$), and the continuity of $dP_{cold}/d\eta$ at $\eta=1$ where $P_{cold}$ must match for compressed and non-compressed states.  Various exponents $1.2 < a < 3.0$ can be substituted into the equation \citep{MeloshMANEOS}, and affect the behavior of the interatomic potential at large distances.  \citet{MeloshMANEOS} advocates a value of $a=1.7$ for quartz, but $a=1.27$ is used here for consistency with the form of the equation of state used {\Revision by \citet{Canup2004}}.  A value of $a=4/3$ would be appropriate for a solid bound by Coulomb forces, $a=7/3$ corresponds to a solid bound by Van Der Waals forces, and also the well-known Lennard-Jones 6-12 potential.  Different values of $a$ shift the critical point and affect the slope of the liquid/vapor phase curve.

The second group of changes involve changes to the Helmholtz free energy due to the different partitioning of energy in a molecular gas versus a purely atomic gas.  \citet{MeloshMANEOS} includes rotational and vibrational contributions to the free energy, but for consistency with the equation of state used in SPH by \citet{Canup2004},  \citet{m-moon} omitted these terms.  Adding these terms to M-ANEOS for dunite is a subject of future work.

\section{Run119: A Benchmarking Case for Giant Impact Simulations  \label{sec:benchmarking}}
\citet{m-moon} used run119 of \citet{Canup2004} as one of their simulations used to compare results of CTH and SPH.  As such, I suggest that run119 could be used as a standard benchmarking case for other hydrocode simulations, as well.  However, not all of the details required to fully reproduce run119 are provided in prior publications.  

Table \ref{table:ser119} describes the physical parameters used to set up run119 in CTH.  The total mass involved in the collision $M_T=1.02M_E$, $\gamma=0.13$, $v_{imp}=v_{esc}$, an impact angle of 46$^{\circ}$.  The sizes of iron core and dunite mantle yield a composition of roughly 70\% dunite, 30\% iron.  {\Revision The target is 29.7\% iron by mass, and the projectile is 30.2 \% iron by mass.}  Table \ref{table:timelogs} gives the results of the simulation, including the disk mass, angular momentum, mass fraction of iron in the disk, and predicted Moon mass, {\Revision using equation (\ref{eq:ICS97}) with the ICS97 coefficients: $a=1.9$, $b=1.15$, and $c=1.9$}.  These quantities can be used to compare outcomes of the Moon-forming simulation across different hydrocodes.

\clearpage
\clearpage 
\begin{table}[h]
   \centering
     \begin{tabular}{ll} 
\hline
Planet & $H_{sat}/(H_{rot}+H_{sat})$ \\
\hline
Earth & 0.831 \\
Mars & $1.5\times10^{-6}$ \\
Jupiter & 0.0107 \\
Saturn & 0.0142 \\
Uranus & 0.0109 \\
Neptune & 0.0203\\
\hline
   \end{tabular}
   \caption{Ratio of the angular momentum in orbital motion of satellites, $H_{sat}$ to the total system angular momentum, $H_{sat}+H_{rot}$, after \citet{MacDonald1966}.}
   \label{table:ang}
\end{table}

\newpage
\begin{table}[h]
   \centering
     \begin{tabular}{lll} 
\hline
Parameter & Symbol & Value\\
\hline
Thermal conductivity of rock & $k$ & 4 W m$^{-1}$ K$^{-1}$ \\
Magma temperature & T & 2000 K \\
Melt density & $\rho_l$ & 3300 kg m$^{-3}$ \\
Thermal expansion & $\alpha$ & $2\times 10^{-5}$ K$^{-1}$ \\
Melt viscosity & $\eta_l$ & 0.1 Pa s \\
Thermal diffusivity & $\kappa$ & $10^{-6}$ m$^{2}$ s$^{-1}$ \\
Rheologically critical melt fraction & $\phi_{cr}$ & 0.4$^{a}$ \\
Latent heat of silicate & $L$ & $4 \times 10^5$ J kg$^{-1}$ \\
Specific heat of liquid silicate & $C_{p,l}$ & 1200 J kg$^{-1}$ K$^{-1}$\\
\hline
   \end{tabular}
   \caption{Thermal and physical properties describing the lunar parent body and its interior magma ocean.  Values from \citet{ReeseSolomatov2006} unless otherwise noted. $^a$\citet{RennerEPSL}.}
   \label{table:magma}
\end{table}

\begin{table}[h]
   \centering
     \begin{tabular}{lll} 
\hline
Parameter & Meaning & Value \\
\hline
V1 & Number of elements & 1 \\
V2 & Switch for type of equation of state & 4 \\
& V2=4 indicates solid-liquid-gas with ionization & \\
V3 & Reference density & 7.85 g/cm$^3$ \\
V4 & Reference temperature & 0 \\
	& a value of 0 defaults to 0.02567785 eV (298 K) &\\
V5 & Reference pressure & 0 \\
V6 & Reference bulk modulus& $1.45 \times 10^{12}$ dyne/cm$^2$\\
V7 & Gruneisen gamma & 1.69 \\
V8 & Reference Debye temperature & 0.025 eV \\
\hline
V9 & $T_{\Gamma}$ parameter & 0\\
V10 & 3$\times$ the limiting value of Gruneisen gamma & 2/3\\
& for large compressions & \\
V11 & Zero temperature separation energy & $8.2 \times 10^{10}$ erg/g \\
V12 & Melting temperature & 0.15588 eV \\
V13 & Parameter $c_{53}$ for low density& 0 \\
	& modification of the critical point &  \\
V14 & Parameter $c_{54}$ for low density& 0\\
	& modification of the critical point &  \\
V15 & Thermal conductivity coefficient & 0 \\
	& if zero, conduction is not included & \\
V16 & Temperature dependence of thermal & 0\\
	& conduction coefficient 		& \\
\hline
V17 & Minimum solid density & 0 \\
 &Solid-solid phase transition parameters &  \\
V18 & $D1$, density at onset of hppt& 0\\
V19 &$D2$	, density at completion of hppt & 0\\
V20 & $D3$, pressure at center of hppt & 0 \\
V21 & $D4$, $dP/dh$ at end of hppt& 0\\
V22 & $D5$, $d^2P/dh^2$ & 0 \\
V23 & Heat of fusion for melting & 2.471$\times 10^9$ erg/g \\
V24 & Ratio of liquid to solid density & 0.955 \\
\hline
V25 -- V48 & Not used in this EOS type & 0 \\
\hline
   \end{tabular}
   \caption{ANEOS coefficients for iron, after \citet{Canup2004}.}
   \label{table:aneos-iron}
\end{table}

\begin{table}[h]
   \centering
     \begin{tabular}{lll} 
\hline
Parameter & Meaning & Value \\
\hline
V1 & Number of elements & 3 \\
V2 & Switch for type of equation of state & 4 \\
& V2=4 indicates solid-liquid-gas with ionization & \\
V3 & Reference density & 3.32 g/cm$^3$ \\
V4 & Reference temperature & 0 \\
	& a value of 0 defaults to 0.02567785 eV (298 K) &\\
V5 & Reference pressure & 0 \\
V6 & Reference sound speed & $-6.6 \times 10^5$ cm/s \\
& in linear shock-particle velocity relationship & \\
	& negative sign required to flag ANEOS & \\
V7 & Gruneisen gamma & 0.82 \\
V8 & Reference Debye temperature & 0.057 eV \\
\hline
V9 & Constant in linear Hugoniot shock-particle & 0.86 \\
& velocity relationship & \\
V10 & 3$\times$ the limiting value of Gruneisen gamma & 2\\
& for large compressions & \\
V11 & Zero temperature separation energy & $1.3 \times 10^{11}$ erg/g \\
V12 & Melting temperature & 0.19 eV \\
V13 & Parameter $c_{53}$ for low density& 1.97$\times 10^{11}$ \\
	& modification of the critical point &  \\
V14 & Parameter $c_{54}$ for low density& 0.8 \\
	& modification of the critical point &  \\
V15 & Thermal conductivity coefficient & 0 \\
	& if zero, conduction is not included & \\
V16 & Temperature dependence of thermal & 0\\
	& conduction coefficient 		& \\
\hline
V17 & Minimum solid density & 0 \\
 &Solid-solid phase transition parameters &  \\
V18 & $D1$, density at onset of high pressure phase transition (hppt)& 4.65 \\
V19 &$D2$	, density at completion of hppt & 4.9 \\
V20 & $D3$, pressure at center of hppt & $6.6 \times 10^{11}$ \\
V21 & $D4$, $dP/dh$ at end of hppt& $3.5 \times 10^{12}$ \\
V22 & $D5$, $d^2P/dh^2$ & $1.3 \times 10^{13}$ \\
V23 & Heat of fusion for melting & 0 \\
	& V23=0 because melting in not included & \\
V24 & Ratio of liquid to solid density & 0.95 \\
	& at the melting point & \\
\hline
V25 -- V32 & Not used in this EOS type & 0 \\
\hline
   \end{tabular}
   \caption{ANEOS coefficients for dunite (forsterite, Mg$_2$SiO$_4$), after \citet{BenzMoonPaper3, m-moon}.  Values
   continue in Table \ref{table:aneos-dunite2}.}
   \label{table:aneos-dunite}
\end{table}

\begin{table}[h]
   \centering
     \begin{tabular}{lll} 
\hline
Parameter & Meaning & Value \\
\hline
V33 & Flag for ionization model & 1 \\
	& 0=Saha, 1=Thomas-Fermi & \\
V34 & $E_{shift}$, shift energy for reactive chemistry & 0 \\
V35 & $S_{shift}$, shift entropy for reactive chemistry & 0 \\
V36 & Number of atoms in molecular clusters & 2 \\
V37 & $E_{bind}$ & 8.0 eV \\
V38 & Rotational degrees of freedom & 0 \\
V39 & $R_{bond}$, length of molecular bond & $1.5\times10^{-8}$ cm\\
V40 & Vibrational degrees of freedom & 0 \\
\hline
V41 & Vibrational Debye temperature & 0.1723\\
V42 & Flags use of Lennard-Jones & 1 \\
  & (in this case, Mie) potential & \\
 V43 & Power in Mie potential & 1.27 \\
 V44-V48 & Not used & \\
 \hline
 Z	&  Atomic numbers for each element & \\
 \hline
 O	& 			& 8 \\
 Mg	& 			&12 \\
 Si	& 			& 14 \\
 \hline
$f$ &	Abundance by number for each element& \\
\hline
 O	& 			& 0.571 \\
 Mg	& 			&0.286 \\
 Si	& 			& 0.143 \\
 \hline
 
   \end{tabular}
   \caption{Table \ref{table:aneos-dunite}, continued.  ANEOS coefficients for dunite (forsterite, Mg$_2$SiO$_4$), after \citet{BenzMoonPaper3, m-moon}. }
   \label{table:aneos-dunite2}
\end{table}

\clearpage
\begin{table}[h]
   \centering
     \begin{tabular}{ll} 
\hline
Parameter & Value \\
\hline
Target radius & 6295.1 km \\
Radius of target iron core & 3155.1 km\\
Target center ($x,y,z$) & (-1142.8,-991.941, 0) km \\
Target $v_x$ & \textbf{1.08474 km s$^{-1}$} \\
Guess for surface temperature & 2000 K \\
Guess for central temperature & 4000 K \\
\hline
Impactor radius & 3587.2 km \\
Radius of impactor iron core & 1892.0 km \\
Impactor center ($x,y,z$) & (7648.7, 6638.9, 0) km \\
Impactor $v_x$ & \textbf{-7.26 km s$^{-1}$} \\
Guess for surface temperature & 2750 K \\
Guess for central temperature & 1700 K \\
\hline
Domain minimum/maximum in $x, y, z$ & $-8 \times 10^5$ km to $8\times 10^5$ km\\
Minimum density for AMR refinement$^a$ & $10^{-2}$ g cm$^{-3}$ \\
\hline
   \end{tabular}
   \caption{Parameters used to simulate run119 in CTH, from \citet{m-moon}. Equation of state parameters used for iron and dunite in run119 are summarized in Appendix \ref{sec:aneos-appendix}.  $^a$For densities above this value, the value of ``maxl,'' the maximum level of mesh refinement in CTH, increases by 1, as the density decreases by a factor of 8.  This yields CTH elements that contain roughly equal mass, similar to the way that each SPH particle contains an equal amount of mass.   \label{table:ser119}}
\end{table}

\begin{table}[h]
   \centering
     \begin{tabular}{llllll} 
\hline
Time (hours) & Disk Mass ($M_L$) & Disk L ($L_{EM}$) & $m_{fe,disk}$  & Predicted Moon Mass ($M_L$)\\
\hline
0 & 4.867 & 0.592 & 0.282& 1.069 \\
1 & 3.698 & 0.475 & 0.200& 1.051 \\
2 & 3.474 & 0.513 & 0.119 & 1.687 \\
3 & 2.901 & 0.492 & 0.103 & 2.063\\ 
4 & 2.884 & 0.511 & 0.112 & 2.281\\
5 & 3.131 & 0.525 & 0.126 & 2.170 \\
6 & 2.867 & 0.524 & 0.162 & 2.441\\
7 & 2.868 & 0.497 & 0.118 & 2.152\\
8 & 2.490 & 0.463 & 0.080 &2.188\\
9 & 2.659 & 0.470 & 0.181 & 2.088 \\
10 & 2.068 & 0.388 & 0.200 &1.848\\
11 & 1.975 & 0.373 & 0.095& 1.784\\
12 &1.777 & 0.349 & 0.091 & 1.735\\
13 & 1.653 & 0.333 & 0.083 & 1.699 \\
14& 1.578 & 0.322 & 0.063 & 1.663\\
15 & 1.564 & 0.318 & 0.057 & 1.645 \\
16 & 1.529 & 0.313 & 0.049 & 1.623 \\
17 & 1.506 &  0.308 & 0.048 & 1.602 \\
18 & 1.490 & 0.305& 0.047 & 1.586 \\
19 & 1.474 & 0.303 & 0.045 & 1.575\\
20 & 1.451 & 0.296 & 0.044 & 1.536 \\
25 & 1.391 & 0.277& 0.040 & 1.399\\
30 & 1.376 & 0.276 & 0.033 & 1.404 \\
35 & 1.333 & 0.268 & 0.291 & 1.363 \\
40 & 1.257 & 0.248 & 0.028 & 1.236 \\
45 &1.158 & 0.221 & 0.028 & 1.059 \\
50 &1.134 & 0.217 & 0.028 & 1.037\\
\hline
   \end{tabular}
   \caption{Evolution of disk mass (scaled by lunar masses), disk angular momentum, mass fraction of iron in the disk, and predicted Moon mass (using {\Revision equation \ref{eq:ICS97} with the ICS97 coefficients}), scaled by lunar masses) for run119.  These data were used to create Figure \ref{fig:ser119-timelogs}.  \label{table:timelogs}}
\end{table}

\clearpage


\newpage

\newpage

\newpage

\newpage

\newpage

\newpage

\newpage

\newpage

\end{document}